\documentclass[journal,12pt,onecolumn,draftclsnofoot,]{IEEEtran}
\IEEEoverridecommandlockouts
\def\BibTeX{{\rm B\kern-.05em{\sc i\kern-.025em b}\kern-.08em
    T\kern-.1667em\lower.7ex\hbox{E}\kern-.125emX}}
    
\usepackage{booktabs} 
\usepackage{acronym}
\usepackage{algorithmic}
\usepackage{algorithm}
\usepackage{enumitem}
\usepackage{bbm}
\usepackage{balance}
\usepackage{url}
\usepackage{subfig}
\usepackage{amsmath, amsthm, amssymb, amsfonts}
\newtheorem{Theorem}{Theorem}[section]
\numberwithin{Theorem}{section}

\usepackage{tikz}
\usetikzlibrary{chains,positioning,matrix,arrows,scopes,shapes,fit,calc}

\acrodef{PRIMA}{PRIvacy Masking and Anonymization}
\acrodef{GDPR}{EU General Data Protection Regulation}

\newcommand{\hbc}{Honest-but-Curious~}

\newcommand{\DO}{DO~}
\newcommand{\CSP}{CSP~}
\newcommand{\FHE}{FHE}
\newcommand{\PHE}{PHE}
\newcommand{\SWHE}{SHE~}
\newcommand{\CA}{$P1~$}
\newcommand{\CB}{$P2~$}

\DeclareMathOperator{\Laplace}{Lap}
\DeclareMathOperator{\Uniform}{Unif}
\DeclareMathOperator{\sgn}{sgn}

\begin{document}

\title{Secure k-Anonymization over Encrypted Databases}

\author{

\IEEEauthorblockN{Manish Kesarwani,}
\IEEEauthorblockA{\textit{IBM Research, India,} 
\textit{manishkesarwani@in.ibm.com}} 
\and

\IEEEauthorblockN{Akshar Kaul},
\IEEEauthorblockA{\textit{IBM Research, India,} 
\textit{akshar.kaul@in.ibm.com}}
\and

\IEEEauthorblockN{Stefano Braghin},
\IEEEauthorblockA{\textit{IBM Research Europe — Ireland},
\textit{stefanob@ie.ibm.com}}
\and

\IEEEauthorblockN{Naoise Holohan},
\IEEEauthorblockA{\textit{IBM Research Europe — Ireland},
\textit{naoise.holohan@ibm.com}}
\and

\IEEEauthorblockN{Spiros Antonatos},
\IEEEauthorblockA{\textit{IBM Research Europe — Ireland},
\textit{santonat@ie.ibm.com}}

}




\maketitle

\begin{abstract}
Data protection algorithms are becoming increasingly important to support modern business needs for facilitating data sharing and data monetization. 
Anonymization is an important step before data sharing. Several organizations leverage on third parties for storing and managing data.
However, third parties are often not trusted to store plaintext personal and sensitive data; data encryption is widely adopted to protect against intentional 
and unintentional attempts to read personal/sensitive data. Traditional encryption schemes do not support operations over the ciphertexts and thus 
anonymizing encrypted datasets is not feasible with current approaches. This paper explores the feasibility and depth of implementing a privacy-preserving
data publishing workflow over encrypted datasets leveraging on homomorphic encryption. We demonstrate how we can achieve uniqueness discovery, data masking,
differential privacy and k-anonymity over encrypted data requiring zero knowledge about the original values. We prove that the security protocols followed by our approach
provide strong guarantees against inference attacks. Finally, we experimentally demonstrate the performance of our data publishing workflow components.
\end{abstract}
\section{Introduction}
\label{sec:intro}

Nowadays, applications interact with a plethora of potentially sensitive information from multiple sources. As an example, modern applications regularly combine data from different domains such as healthcare and IoT. While such rich sources of data are extremely valuable for analysts, researchers, marketers and other professionals, data privacy technologies and practices face several key challenges to keep pace.

There are two major obstacles when it comes to hosting and sharing sensitive data. The first is that the public cloud solutions are not trusted with sensitive data (e.g.\ health, financial, or critical infrastructure data) and thus organisations have to invest in private or hybrid clouds as
the hosting and processing environment. This adds complexity for the design and implementation and often comes with additional cost due to security and customisation. Homomorphic encryption provides an answer to these kinds of obstacles, by encrypting the data at their source while allowing operations on them and thus lifting the trust barrier from the hosting solution.

The second is data privacy. Data privacy technologies are applied in two major use cases. The first use case concerns 
data sharing, where data need to be sufficiently anonymized before being shared with researchers and analysts. The second
use case concerns security. By anonymizing data at rest, the risk of breaches is minimised since sensitive information
is protected. 

Latest advances in regulation, like \ac{GDPR},
 also propose anonymization for safely processing data when consent
is not an option or organisations want to use them for purposes beyond those for which it was originally obtained for an indefinite
period of time.

In this paper, we present how to apply different data privacy approaches to homomorphically encrypted data. Specifically,
we present how we can achieve uniqueness discovery, data masking,
differential privacy and $k$-anonymity over encrypted data, requiring zero knowledge about the original values. Uniqueness discovery
allows the user to find which attributes or combinations of attributes (quasi-identifiers) appear with a lower frequency than a
predefined threshold. This leads to the selection of attributes that need to be protected via a combination of data masking, differential privacy
and $k$-anonymity approaches. We explore how we can securely apply all these techniques without leaking information about the original data,
such as the domain cardinality or diameter.

The rest of the paper is organised as follows. Section~\ref{sec:motivation} describes the motivating scenarios and use cases
behind this work. Section~\ref{sec:background} provides background information about the basic
principles of data privacy and homomorphic computations and further outlines the data publishing workflow and describes the building blocks to achieving data privacy. 
In Section~\ref{sec:related} we present the related work.
In Section~\ref{sec:protocols} we present the secure protocols while in Section~\ref{sec:security} and Section~\ref{sec:performance}
we discuss their security guarantees and performance respectively. Finally, we conclude in Section~\ref{sec:conclusions}.

\section{Motivation}
\label{sec:motivation}

The first major question that arises is why data encryption alone is not enough. In general, to protect the privacy of sensitive data, only encrypted data is outsourced to the third-party cloud providers and there exist well-established systems which allow secure query processing directly over encrypted data.
On top of that, a more important question is, why someone should perform masking and anonymization over encrypted data in the cloud environment. In a typical scenario, the data is anonymized at the source and then uploaded to the cloud or shared with a third party. However, there are many reasons to perform the anonymization part on the cloud after data encryption. 

The answer to these two questions relies on two major observations from the GDPR compliance standard. First, data encryption does not meet the high compliance standards, since all the data encryption schemes are reversible. 
Second, anonymized data is not considered personal information and benefit from relaxed standards under GDPR\footnote{\url{https://ibm.biz/Bd2yUK}}.
Thus the need to apply non-reversible masking and anonymization over encrypted data is required. 
Furthermore, the GDPR mandates that the data controller needs to demonstrate that the state-of-the-art strategy is applied when it comes to pseudonymization/anonymization approaches. 
By relying on the cloud to deploy the state-of-the-art approaches, the operational and compliance model for the data owners becomes significantly easier. 

Apart from the compliance regulations, there are several other factors that motivate anonymization over encrypted data.
First, the objective of the data use may change over the course of time. Initially, it may not be desirable to share the data but after some time
it might be required. This is common with enterprise data where confidentiality must be kept for several years before the data can be exchanged. 
Second, the users might want to selectively share data and thus apply anonymization to only the selected portion. In both cases,
by pre-anonymizing the data we will not be able to reach the desired outcome. Furthermore, in the distributed IoT scenario, individual sensors may not have sufficient storage and processing capability and the ever-increasing volume of data makes it hard to anonymize at the source. As an operating model, it is far easier to homomorphically encrypt the data at source before outsourcing to the cloud and then anonymize on-demand at the cloud when it comes to sharing and collaborating. 

Specific masking operations, k-anonymity and differential privacy fall into the non-reversible anonymization category and thus are the focus of this paper.

Our approach guarantees two major properties: a) the data owner does not send the data as-is and so the data trust cannot see the original data and 
b) facilitate the application of on-demand non-reversible anonymization approaches to the data in order to meet compliance standards or selective data sharing.

\section{Background}
\label{sec:background}

\subsection{$k$-Anonymity}

Based on the data privacy terminology, attributes in a dataset are classified as direct or quasi identifiers.
Direct identifiers are uniquely identifying and are always removed or masked before the data release.
Quasi-identifiers are sets of attributes that can uniquely identify one or very few individuals.
For example, for the dataset in Table \ref{tbl:quasi}, if we observe the gender attribute in isolation, it is not uniquely identifying (roughly 50\% of a dataset
would be either male or female); the same applies for a ZIP code (several thousands of people might live in the same ZIP code).
However, if we look at attributes in combinations then we can isolate very few individuals. As an example,
the combination of ZIP code plus gender plus birth date can be uniquely identifying (in case of the US this combination
can uniquely identify 87$\%$ of the population). 
Based on the $k$-anonymity approach~\cite{sweeneykanon}, quasi-identifiers are generalized and clustered in such a degree that an individual is indistinguishable from at least $k-1$ other individuals.

\begin{table}
\centering
\small
    \caption{Example dataset with one direct identifier (name) and age plus gender plus ZIP code as quasi-identifiers}
    \label{tbl:quasi}
    \begin{tabular}{|l|l|l|l|l|}
        \hline
                \textbf{Record ID} & \textbf{Name} & \textbf{Age} & \textbf{Gender} & \textbf{ZIP} \\
        \hline
                \hline
        1 & John & 18 & Male & 13122 \\
        2 & Peter & 18 & Male & 13122 \\
        3 & Mark & 19 & Male & 13122 \\
        4 & Steven & 19 & Male & 13122 \\
        5 & Jack & 18 & Male & 13121 \\
        6 & Paul & 20 & Male & 13121 \\
        7 & Andrew & 20 & Male & 13121 \\
      \hline
    \end{tabular}

\end{table}

\subsection{System Entities}
\label{subsec:entities}
A Database-as-a-Service (DBaaS) architecture consists of below entities:
\begin{enumerate}

\item {\bf Data Owner (DO)}: A company or an individual who is having a proprietary right to a sensitive database, such as a Bank.
The Data Owner wants to securely outsource its data storage and future computations over data to a Cloud Service Provider.

\item {\bf Cloud Service Provider (CSP)}: A third party, that provides the storage and computation capability as a service to its clients. For our scenario, the Cloud Service Provider is a system where any present-day state of the art database engine is running. For example, Amazon Redshift and Microsoft Azure SQL Database.

In particular, we work in the two-party federated cloud setting, with two non-colluding public cloud servers. This model was introduced in Twin Clouds~\cite{twincloud} and was subsequently used in related problems \cite{secureKNN, manish}. Federated clouds are an example of Interclouds~\cite{grozev2014inter}, a collection of global stand-alone clouds. Intercloud allows better load balancing and allocation of resources.  A detailed survey of the taxonomy of intercloud architectures is presented in~\cite{grozev2014inter}. 

\end{enumerate}

\subsection{Trust Assumption}
\label{subsec:trust}
We assume that the \CSP is honest-but-curious (or semi-honest) i.e. it is honest and executes the protocol correctly, but is also interested in the plaintext of the encrypted data stored at its site, either because it is curious or it has been compromised. In this paper, we will show that the honest-but-curious adversary will not be able to learn anything about the plaintext of the encrypted database, even though it can observe the computations and can take memory dumps. Further, we also prevent the leakage of any data clustering information available in the intermediate steps of secure $k$-anonymization, masking or differential privacy algorithms.

\begin{figure*}[ht!]
  \centering
  \includegraphics[width=0.7\textwidth]{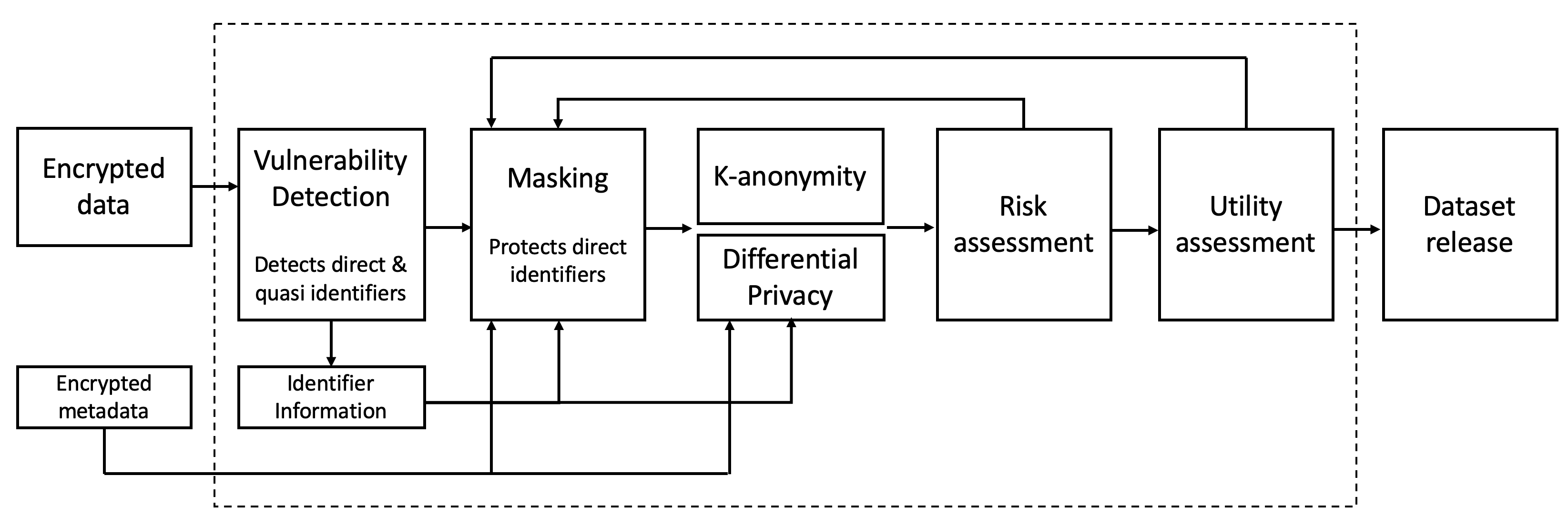}
  \caption{The workflow for encrypted data de-identification process}
  \label{fig:workflow}
\end{figure*}

\subsection{Homomorphic Computation}
\label{subsec:swhe}

%

Homomorphic encryption schemes support direct computation of functions over encrypted data without needing to decrypt it first. To this end the seminal work of Gentry \cite{fhe} presents a fully homomorphic encryption (\FHE) scheme, which is capable of evaluating any arbitrary dynamically chosen function over an encrypted database without needing the secret key. But since computation over fully homomorphic encrypted data is still many orders of magnitude slower than the plaintext execution, this limits the practical deployment of these scheme for real workloads.

Another line of research built partial and somewhat homomorphic encryption schemes. Specifically partial homomorphic encryption (\PHE) schemes support evaluation of a chosen atomic function over encrypted data (like Addition or Multiplication). For example, Paillier cryptosystem \cite{paillier} supports addition over encrypted data without needing a secret key and ensures strong security guarantees. On the other hand, the somewhat homomorphic encryption (\SWHE\!\!) scheme supports the computation of low degree polynomials over encrypted data. For example, BGN cryptosystem \cite{bgn} supports evaluation of any polynomial of degree \emph{two} over encrypted data, while in LFHE cryptosystem \cite{lfhe}, degree \emph{d} polynomials can be evaluated, but it bases security on weaker assumptions of learning with error (LWE) or ring-LWE (RLWE) problems.


In general, a \SWHE~ encryption scheme consists of five basic algorithms: 
a) key generation $SHE.KeyGen(1^{\lambda})$ that takes as input the security parameter $\lambda$ and output the secret key $sk$, the public key $pk$ and public parameters $params$ 
b) encryption $SHE.Enc(params, pk, m)$ that takes the message $m$ and evaluates the corresponding ciphertext $c$ using $params$ and the public key $pk$ 
c) decryption $SHE.Dec(params, sk, c)$ that takes the ciphertext $c$ and decrypts it using $params$ and the secret key $sk$ and outputs the corresponding plaintext message $m$
d) addition $SHE.Add(pk,c_1,c_2)$ which takes two ciphertexts $c_1$ and $c_2$ and adds them homomorphically such that the output $c_3 \leftarrow FHE.Enc(params, pk, m_1+m_2)$, where $m_1$ and $m_2$ are the plaintext corresponding to the input ciphertexts $c_1$ and $c_2$ respectively and finally
e) multiplication $SHE.Mult(pk,c_1,c_2)$ which multiplies  homomorphically two ciphertexts $c_1$ and $c_2$ such that the output $c_3 \leftarrow FHE.Enc(params, pk, m_1*m_2)$.

%
%
%
%
%
%
%

In this paper, we use the LFHE encryption scheme to compute squared Euclidean distance over encrypted databases, which is a key building block of our secure anonymization protocol. The details of LFHE cryptosystem can be found at~\cite{lfhe}.

\subsection{Differential Privacy}\label{sc:bg:dp}
Differential privacy is a more recent development in the field of privacy-preserving data publishing and data mining.
Achieved by adding randomness to the data, differential privacy renders individuals' data and data mining outputs statistically indistinguishable, thereby protecting individuals' privacy~\cite{Dwo06}.
Differential privacy has been shown to provide strong guarantees against auxiliary information attacks~\cite{KS08,GKS08}, and in recent years has been adopted by large corporations when collecting\slash publishing sensitive data~\cite{CK12,App16,EPK14}.

A mechanism M is said to be $\epsilon$-differentially private if adding or removing a single data item in a database only affects the probability of any outcome within a small multiplicative factor.
 Formally, a randomized mechanism M is $\epsilon$-differentially private if for all data sets $D_1$ and $D_2$ differing on at most one element, and all $S \subseteq Range(M) $ then
  $$ Pr[M(D_1) \in S] \leq exp({\epsilon}) \cdot Pr[M(D_2) \in S]$$

There is a number of mechanisms available to achieve local differential privacy~\cite{NIPS2014_5392}, covering many different types of data.
When working with continuous numerical data, differential privacy is commonly achieved using the Laplace mechanism~\cite{DMN06}.
The authors showed that, by adding noise from a suitably-scaled Laplace distribution, the resulting output will satisfy differential privacy.
The geometric mechanism is a discrete variant of the Laplace mechanism, used when dealing with integer-valued data~\cite{GRS12}.
%
For binary-valued data, differential privacy can be achieved by flipping values at random. The probability for flipping is equal to $\frac{1}{e^{\epsilon} + 1}$~\cite{HLM17}.
In some cases, the addition of noise to the data does not make sense, e.g.\ categorical data.
In this context, the exponential mechanism provides a means to achieve differential privacy.
Developed by McSherry and Talwar~\cite{MT07}, the exponential mechanism selects an output at random, weighted by a utility function which is specified by the data controller.
%


\subsection{Our Contributions}
\label{subsec:contribution}

In this work we build a secure privacy-preserving data publishing workflow over encrypted datasets. The workflow consists of five major components.
Figure~\ref{fig:workflow} illustrates the steps of the workflow. The workflow expects as input the encrypted data as well as encrypted meta-data, such as the dictionaries to be used for masking
or encrypted parameters for differential privacy. In Section~\ref{sec:protocols} we present the details of our secure protocols.

\begin{enumerate}
  \item \textbf{Secure Privacy Vulnerability Identification}. This step detects direct identifiers and combinations of attributes-values (quasi-identifiers)~\cite{sweeneykanon} that lead to high re-identification risk. The detection is based on the attribute values. In Section~\ref{subsec:pvi} we illustrate how direct and quasi-identifiers can be obtained from an encrypted database.

  \item \textbf{Secure Data Masking}. This component protects the direct identifiers detected by the privacy vulnerability identification component.
    The values of direct identifiers can be replaced with fictionalized values or redacted; the action taken is based on a pre-defined configuration.
    
  \item \textbf{Secure k-anonymity and differential privacy}. This component protects the quasi-identifiers, by applying algorithms with strong security guarantees, such as  differential privacy and $k$-anonymity. Here data are generalized and\slash or suppressed and\slash or perturbed so that the re-identification risk becomes smaller than a pre-specified threshold. 
  
  \item \textbf{Risk assessment}. This component assesses the risk associated with the dataset. It is an additional step during exploratory phases in which expert assessors and policymakers are still evaluating what additional privacy constraints to apply, in addition to what is required by the current legislation.
    
  \item \textbf{Utility assessment}. This component allows the estimation of the loss in utility caused by the de-identification\slash anonymization process.
\end{enumerate}

%
%
%
%
%


\section{Related work}
\label{sec:related}

L. Sweeney et al.~\cite{sweeneykanon} introduced the concept of k-anonymity and how it can protect against re-identification
attacks via creating indistinguishable records. Khaled El Emam et al.~\cite{ola} proposed a way to achieve globally optimal
$k$-anonymity. LeFevre et al.~\cite{mondrian} proposed Mondrian as an approach to achieve good balance between generalization and information 
loss for multidimensional datasets. These works, along with numerous others that present optimal solution to achieve $k$-anonymity,
try to prevent re-identification attacks through generalization. All these approaches work on unmodified data and they do not include the notion
of anonymity over encrypted datasets.

Achieving $k$-anonymity using clustering is not a new concept. Bertino et al.~\cite{bertinokmember} proposed an efficient k-anonymization algorithm
called $k$-member, which is useful in identifying required generalization to apply $k$-anonymity to a given dataset. Loukides and Shao~\cite{loukidesclustering}
propose novel clustering criteria that treat utility and privacy on equal terms and propose sampling-based techniques to optimally set up its parameters.
Aggarwal et al.~\cite{aggarwalclustering} use a personalized clustering algorithm in order to provide a level of anonymity to the individuals recorded in the dataset.
All of the proposed algorithms require direct access to the data and do not operate over encrypted data.

Jiang and Clifton~\cite{jiangkanon} propose a secure distributed framework for achieving $k$-anonymity. 
Their paper describes a method to locally anonymize dataset so that the joined dataset will be $k$-anonymous.
A two-party secure distributed framework is developed which can be adopted to design a secure protocol to compute $k$-anonymous data from two vertically partitioned sources.
This framework does not apply to encrypted data shared in an hybrid cloud infrastructure. Jiang and Atzori~\cite{atzorikanon} 
propose a privacy-preserving strategy to mine $k$-anonymous frequent item sets between two, or more, parties.
The proposed algorithm operate on encrypted data to extract insights. The original data are not modified and they are still not compliant with any privacy model 
after the application of the proposed algorithm.


Differential privacy and homomorphic encryption has been considered previously.
In~\cite{10.1007/978-3-319-69659-1_17}, \emph{differentially private encryption schemes} were considered as a way to prevent leakage of information.
The authors proposed the Encrypt+DP concept, that imposes differential privacy on the decryption process, rendering it a stochastic process that not always be correct.
They also propose DP-then-Encrypt, whereby noise satisfying differential privacy is first added to the data before being encrypted.
Both of these schemes are different from the one presented in this paper, as we achieve differential privacy on encrypted data, without having to see the plaintext and without having to decrypt the ciphertext.

The work that is closely related to our paper is the approach proposed by Liu et al.~\cite{liuhomomorphic}.
In this paper a method for performing $k$-means over homomorphic encrypted data is presented. The paper uses a specific encryption scheme.
The main difference is that their approach does not extend to $k$-anonymity and that the execution scenario described in their approach 
assumes that the clustering algorithm is performed in a single VM. Furthermore, our work extends to vulnerability identification, masking and differential privacy.

\ac{PRIMA}~\cite{prima} provides several features for the strategy design and enforcement of data privacy in production grade systems.
\ac{PRIMA} aims to guide decision makers through the data de-identification process while minimizing required input.
\ac{PRIMA} operates on a different trust model, where the data are anonymized before reaching or in the cloud environment, 
and has no ability to work on encrypted data.


\section{Secure Protocols}
\label{sec:protocols}

\begin{figure}[tb]
    \centering
      \includegraphics[width=.9\linewidth, height=3cm]{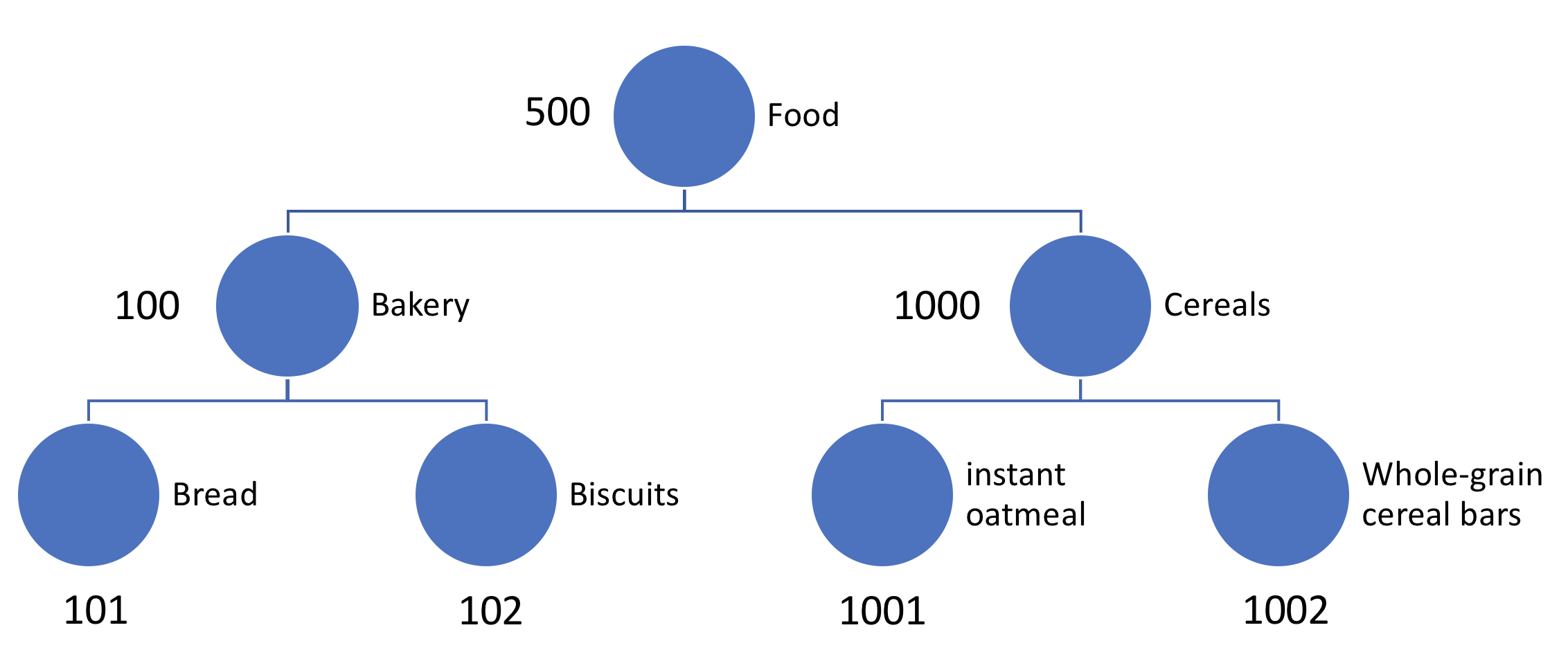}
     \caption{Categorical attribute assignment}
     \label{fig:cat}
  \end{figure}

As described in Section \ref{subsec:entities}, all the protocols proposed in the paper are considered in the two-party \hbc cloud setting. 
The Data Owner (\emph{DO}) has a plaintext database table $\mathcal{T}$ consisting of \emph{N} data points $\{t_1, t_2, \cdots, t_N\}$. Each data point is a $d$ dimensional value, i.e. $t_i = \{t_i^1, t_i^2, \cdots, t_i^d\}$. Furthermore, let the domain of plaintext space be $\mathcal{P}$ and the domain of ciphertext space be $\mathcal{R}$.
The \emph{DO} calls the \emph{KeyGen} function of the \SWHE algorithm to get the public key and secret key pair (\emph{pk, sk}). Next, the \emph{DO} encrypts the plaintext database $\mathcal{T}$ using \emph{pk} to generate an encrypted database $\mathcal{T}^*$ such that $t_i^* = \{Enc_{pk}(t_i^1), Enc_{pk}(t_i^2), \cdots, Enc_{pk}(t_i^d)\}$.
Please note, before encryption all the decimal values are converted to nearest integer. 
Further, the categorical values are first divided into different hierarchy levels from general to specific and each separate path in the hierarchy is assigned values from far-apart ranges as shown in Figure \ref{fig:cat}. 
These assigned values are then considered as representatives for categorical data. The values of the hierarchy are also encrypted on the \emph{DO} side.
This specific assignment technique will help us to securely identify common ancestor as shown in Section~\ref{subsubsec:dataanon}.

Then, the \emph{DO} shares \emph{pk}, $\mathcal{T}^*$ and the identification threshold $k$ with Party \CA and \emph{sk} with Party \CB.

\subsection{Privacy Vulnerability Identification}
\label{subsec:pvi}

The privacy vulnerability identification process explores the combinatorial space of data attributes and aims to identify direct and quasi-identifiers --
value sets that appear fewer times than a pre-defined identification threshold $k$. 
The process starts by inspecting single attributes and tries to find values that appear fewer than $k$ times.
All attributes detected to have values appearing fewer than $k$ times are reported as direct identifiers.
In our example, each name value appears only once, thus the name attribute is a direct identifier.
The process then starts to inspect pairs of attributes that are not direct identifiers, then the algorithm proceeds to inspect combinations of three identifiers and so on.

\begin{figure}[bt]
  \centering
  \scriptsize
    \begin{tikzpicture}
\tikzstyle{every node}=[font=\scriptsize]
    \node(o) {$\varnothing$};
    \node(h1)[below= 3mm of o] {};
    \node(g) [left =.75mm of h1]{$\{G\}$};
    \node(b) [left of=g] {$\{B\}$};
    \node(z) [right =.75mm of h1] {$\{Z\}$};
    \node(m) [right of=z] {$\{M\}$};
    \node(h2) [below of = h1] {};
    \node(bm) [left = -.2mm of h2] {$\{B, M\}$};
    \node(bz) [left of=bm] {$\{B, Z\}$};
    \node(bg) [left of=bz] {$\{B, G\}$};
    \node(gz) [right = -.2mm of h2] {$\{G, Z\}$};
    \node(gm) [right of=gz] {$\{G, M\}$};
    \node(zm) [right of=gm] {$\{Z, M\}$};
    \node(h3) [below of = h2] {};
    \node(bgm) [left = -.2mm of h3] {$\{B,G,M\}$};
    \node(bgz) [left = 1.5mm of bgm] {$\{B,G,Z\}$};
    \node(bzm) [right = -.2mm of h3] {$\{B,Z,M\}$};
    \node(gzm) [right = 1.5mm of bzm] {$\{G,Z,M\}$};
    \node(bgzm) [below = 5mm of h3]{$\{B,G,Z,M\}$};
    \draw(o) -- (g);
    \draw(o) -- (b);
    \draw(o) -- (z);
    \draw(o) -- (m);
    \draw(b) -- (bg);
    \draw(b) -- (bz);
    \draw(b) -- (bm);
    \draw(g) -- (bg);
    \draw(g) -- (gz);
    \draw(g) -- (gm);
    \draw(z) -- (bz);
    \draw(z) -- (gz);
    \draw(z) -- (zm);
    \draw(m) -- (bm);
    \draw(m) -- (gm);
    \draw(m) -- (zm);
    \draw(bg) -- (bgz);
    \draw(bg) -- (bgm);
    \draw(bz) -- (bgz);
    \draw(bz) -- (bzm);
    \draw(gz) -- (bgz);
    \draw(gz) -- (gzm);
    \draw(gm) -- (bgm);
    \draw(gm) -- (gzm);
    \draw(zm) -- (bzm);
    \draw(zm) -- (gzm);
    \draw(bgz) -- (bgzm);
    \draw(bgm) -- (bgzm);
    \draw(bzm) -- (bgzm);
    \draw(gzm) -- (bgzm);
    \node(o1) [left = 20mm of o]{};
    \path[->, draw, color=red, line width=1pt] (o1) to[in=180, out= -90] (b);
    \path[draw, color=red, line width=1pt] (b) to[in=180,out=0] (g);
    \path[draw, color=red, line width=1pt] (z) to[in=180,out=0] (m);
    \path[->, draw, color=red, line width=1pt] (m)   to[in=140,out=-35] (bg);
    \path[draw, color=red, line width=1pt] (bg) to[in=180,out=0] (bz);
    \path[draw, color=red, line width=1pt] (bz) to[in=180,out=0] (bm);
    \path[draw, color=red, line width=1pt] (bm) to[in=180,out=0] (gz);
    \path[draw, color=red, line width=1pt] (gz) to[in=180,out=0] (gm);
    \path[draw, color=red, line width=1pt] (gm) to[in=180,out=0] (zm);
    \path[->, draw, color=red, line width=1pt] (zm)  to[in=140,out=-35] (bgz);
    \path[draw, color=red, line width=1pt] (bgz) to[in=180,out=0] (bgm);
    \path[draw, color=red, line width=1pt] (bgm) to[in=180,out=0] (bzm);
    \path[draw, color=red, line width=1pt] (bzm) to[in=180,out=0] (gzm);
    \path[->, draw, color=red, line width=1pt] (gzm) to[in=90,out=-35] (bgzm);
    \path[draw, color=red, line width=1pt] (g) to[in=180,out=0] (z);
  \end{tikzpicture}
  \caption{A lattice representing all possible combination of attributes and a possible order of subsets generation}
\label{fig:lattice_toy}
\end{figure}
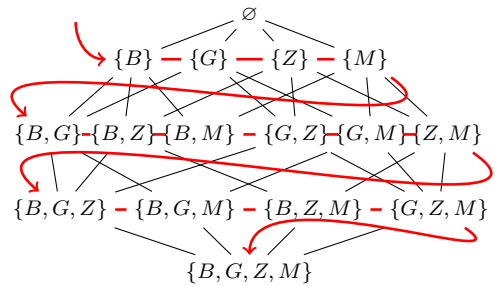

\begin{figure}[bt]
  \centering
  \scriptsize
  \subfloat[Pruning $\{M\}$]{%
    \label{fig:lattice_pruning_m}
    \resizebox{.33\columnwidth}{!}{%
      \begin{tikzpicture}
        \node(o) {$\varnothing$};
        \node(h1)[below  of = o] {};
        \node(g) [left = -1mm of h1]{$\{G\}$};
        \node(b) [left of=g] {$\{B\}$};
        \node(z) [right = -1mm of h1] {$\{Z\}$};
        \node(m) [color=red,right of=z] {$\{M\}$};
        \node(bz) [below of=h1] {$\{B, Z\}$};
        \node(bg) [left of=bz] {$\{B, G\}$};
        \node(gz) [right of= bz] {$\{G, Z\}$};
        \node(bgz) [below of = bz] {$\{B,G,Z\}$};
        \draw(o) -- (g);
        \draw(o) -- (b);
        \draw(o) -- (z);
        \draw(o) -- (m);
        \draw(b) -- (bg);
        \draw(b) -- (bz);
        \draw(g) -- (bg);
        \draw(g) -- (gz);
        \draw(z) -- (bz);
        \draw(z) -- (gz);
        \draw(bg) -- (bgz);
        \draw(bz) -- (bgz);
        \draw(gz) -- (bgz);
      \end{tikzpicture}
    }
  }
  \subfloat[Pruning $\{B,Z\}$]{%
    \label{fig:lattice_pruning_bz}
    \resizebox{.33\columnwidth}{!}{%
      \begin{tikzpicture}
        \node(o) {$\varnothing$};
        \node(h1)[below  of = o] {};
        \node(g) [left = -1mm of h1]{$\{G\}$};
        \node(b) [left of=g] {$\{B\}$};
        \node(z) [right = -1mm of h1] {$\{Z\}$};
        \node(m) [color=red,right of=z] {$\{M\}$};
        \node(bz) [color=red, below of=h1] {$\{B, Z\}$};
        \node(bg) [left of=bz] {$\{B, G\}$};
        \node(gz) [right of= bz] {$\{G, Z\}$};
        \node(bgz) [below = 5mm of bz]{};
        \draw(o) -- (g);
        \draw(o) -- (b);
        \draw(o) -- (z);
        \draw(o) -- (m);
        \draw(b) -- (bg);
        \draw(b) -- (bz);
        \draw(g) -- (bg);
        \draw(g) -- (gz);
        \draw(z) -- (bz);
        \draw(z) -- (gz);
      \end{tikzpicture}
    }
  }
  \subfloat[Pruning $\{G,Z\}$]{%
    \label{fig:lattice_pruning_gz}
    \resizebox{.33\columnwidth}{!}{%
      \begin{tikzpicture}
        \node(o) {$\varnothing$};
        \node(h1)[below  of = o] {};
        \node(g) [left = -1mm of h1]{$\{G\}$};
        \node(b) [left of=g] {$\{B\}$};
        \node(z) [right = -1mm of h1] {$\{Z\}$};
        \node(m) [color=red,right of=z] {$\{M\}$};
        \node(bz) [color=red,below of=h1] {$\{B, Z\}$};
        \node(bg) [left of=bz] {$\{B, G\}$};
        \node(gz) [color=red,right of= bz] {$\{G, Z\}$};
        \node(bgz) [below = 5mm of bz]{};
        \draw(o) -- (g);
        \draw(o) -- (b);
        \draw(o) -- (z);
        \draw(o) -- (m);
        \draw(b) -- (bg);
        \draw(b) -- (bz);
        \draw(g) -- (bg);
        \draw(g) -- (gz);
        \draw(z) -- (bz);
        \draw(z) -- (gz);
      \end{tikzpicture}
    }
  }
  \caption{Impact of pruning to the search space of Figure~\ref{fig:lattice_toy}. Each pruning step is executed after the other.}
  \label{fig:lattice}
\end{figure}
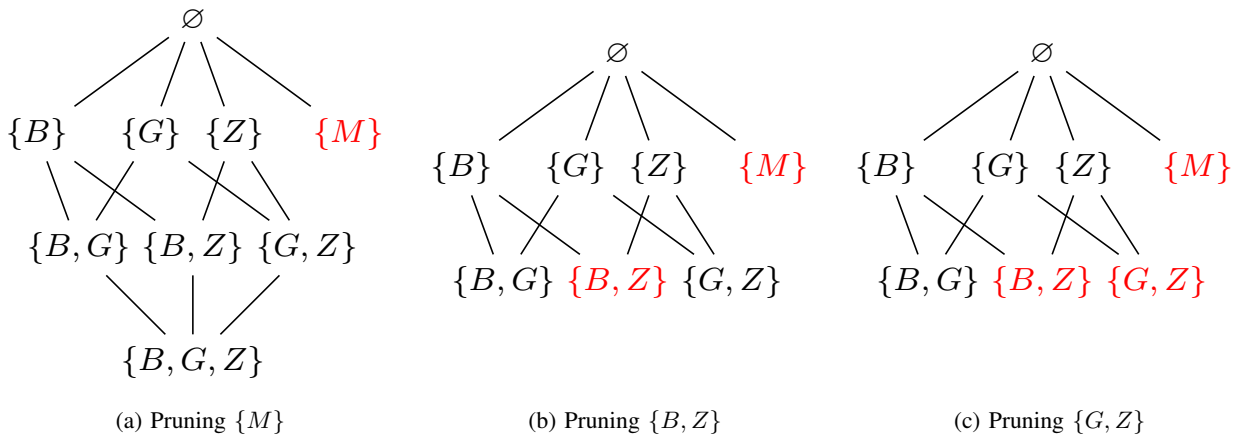

Na\"ive exploration of the entire combinatorial space is infeasible for a large number of attributes since for $d$ attributes $2^d$ combinations need to be checked (see Figure~\ref{fig:lattice_toy}).
Pruning techniques are employed to avoid the exploration of the full space. Pruning can be applied in the following two scenarios.
First, if an attribute, or a set of attributes, $T$ is a quasi-identifier then all the combinations of attributes including this attribute, or set of attributes, are also quasi-identifiers.
Second, if $T$ is not a quasi-identifier then all subset combinations of $T$ are not quasi-identifiers.
This leads to a dramatic reduction of the number of combinations of attributes that need to be checked, thus resulting in a significant improvement in execution time of the protocol.
As an example, consider the scenario shown in Figure~\ref{fig:lattice}. Here the impact of pruning is depicted in terms of the reduction of the search space. Refer to~\cite{fpvi} for further discussion of the impact of pruning in the identification of privacy vulnerabilities.

We use Algorithm~\ref{alg:direct} to identify direct identifiers. Since our encryption function is non-deterministic, a direct comparison of encrypted attribute values will not be helpful. So for each encrypted value in the attribute, \CA computes its difference from the remaining $N-1$ values, where $N$ is the number of tuples in the dataset (the difference will be zero if there is a value match within the attribute). Then \CA multiples these differences with random values in the matrix $\mathcal{R}^*$ and send the computed matrix $\mathcal{M}^*$ to party \CB.
This is shown in Steps $1-7$ of Algorithm~\ref{alg:direct}.
Next, for each attribute, \CB counts the number of zeros for every encrypted value, if there is an encrypted value for which the count of zeros is less than $k$, then this attribute is a direct identifier. \CB tracks the direct identifiers by setting to $1$ the corresponding index in vector $\mathcal{V}$. This is shown in Steps $8-18$ in Algorithm~\ref{alg:direct}. Then, \CB returns the vector $\mathcal{V}$ to Party \CA.

\begin{algorithm}[h]
\small
\caption{\bf \textit{Direct Identifier (DI)}}
\label{alg:direct}
\begin{algorithmic}[1]
\setlength{\itemsep}{-1pt}
\REQUIRE \CA has $\mathcal{T}^*$ and $pk$; \CB has $sk$
\ENSURE \CA learns which attributes of $\mathcal{T}^*$ are direct identifiers
\STATE \CA:
\label{d:a}
\begin{enumerate}
\setlength{\itemsep}{-1pt}
\item {\bf for} $ i = 1 $ to $d$  {\bf do}
    \item \hspace*{10pt} Compute $\mathcal{D}^*_{uv} = t_u^{i*} - t_v^{i*} ~ \forall u,v \in N$
    \item \hspace*{10pt} Sample a matrix $\mathcal{R}$ of size $N \times N ; R_{uv} \in_R \mathcal{P} $
    \item \hspace*{10pt} Compute $\mathcal{R}^* \ni \mathcal{R}^*_{uv} \leftarrow Enc_{pk}(\mathcal{R}_{uv})$
    \item \hspace*{10pt} Evaluate Hadamard product $\mathcal{M}_i^* \leftarrow \mathcal{D}^* \circ \mathcal{R}^*$ 
\item {\bf end for}
\item {\bf Send} $\mathcal{M}^*$ to \CB
\label{d:a:2}
\end{enumerate}
\STATE \CB:
\begin{enumerate}
\setcounter{enumi}{7}
\setlength{\itemsep}{-1pt}
\item Create a vector $\mathcal{V}$ of length $d$
\item Set $\mathcal{V}_i = 0 ~\forall i \in d$
\item {\bf for} $ i = 1 $ to $d$  {\bf do}
    \item \hspace*{10pt} {\bf for} $ j = 1 $ to $N$  {\bf do}
        \item \hspace*{20pt} count = 0
        \item \hspace*{20pt} count += $\mathcal{I}(Dec_{sk}(\mathcal{M}^{*}_{ijl}) == 0)  ~ \forall l \in N$
        \item \hspace*{20pt} {\bf if} count $< k$ {\bf then}
        \item \hspace*{30pt} $\mathcal{V}_i = 1$; break;
        \item \hspace*{20pt}  {\bf end if}
    \item \hspace*{10pt} {\bf end for}
\item {\bf end for}
\item {\bf Return} $\mathcal{V}$ to \CA
\end{enumerate}
\end{algorithmic}
\end{algorithm}

Similarly, we utilize the matrix $\mathcal{M}^*$ computed in Algorithm~\ref{alg:direct} along with the pruning mechanism described earlier to find quasi-identifiers.


\subsection{Data Masking}
\label{subsec:masking}

Data masking is applied when there is need to replace the original values with fictionalized ones.
If we operate on non-encrypted data, then multiple options are available: 
format-preserving and semantic-preserving masking, compound masking as well as some generic masking providers, like nullification, hashing, randomization, truncation and numeric value shifting.
Format-preserving masking dictates that the masked value will have the same format as the original one. 
Semantic-preserving masking ensures that parts of the original value that contain auxiliary information need to be maintained.

Since we operate on encrypted data, not all options are available. 
Semantic-preserving and format-preserving masking cannot be applied since they require access to the original value unless the data owner encrypts only the unique parts of the value. This requires additional metadata so the cloud environment knows how to handle each value (e.g.\ offsets and lengths of encrypted portions of the value). 
However, in this paper, we apply following masking operations:

\begin{itemize}
	\item {\bf Masking of dictionary-based entities}. Entities like names, organization, cities, countries and many more rely on dictionaries to perform format-preserving masking. 
		For example, if we want to replace a name with another one, then we pick a random name from its dictionary. 
		We can apply the same operation over encrypted data. The user uploads a fully encrypted dictionary for the attribute. 
		Then we select a random value from the encrypted dictionary and replace the value. 
		However, the encrypted version of the dictionary needs to be immune to inference attacks. 
		For specific attributes, an attacker can infer the attribute type and values based on cardinality attacks. 
		As an example, a dictionary of two entries could potentially be a gender dictionary. To alleviate this problem, we can append copies of its values to the dictionary.
		Since the encryption is non-deterministic, we can increase the cardinality of the values infinitely. 

	\item {\bf Numerical masking operations}: We can mask numerical values by using the following mechanisms:
		\begin{itemize}
			\item {\bf Add a constant shift amount}, for example adding value 10 to all values
			\item {\bf Noise addition}. Given percentage $x,  0 < x < 1$, we can mask the value $v$ and replace it with a random value in the range $v-v*x$ to $v+v*x$
			\item {\bf Randomization}. Replace a value with a randomly generated number.
		\end{itemize}
	
	\item {\bf Redaction / fixed replacement}: This is a special case where we create dictionaries with encryption of empty string or fixed values.

\end{itemize}

\subsection{Differential Privacy}
\label{subsec:diffpriv}

In this section we demonstrate achieving differential privacy on encrypted data with select mechanisms.
To implement the differential privacy mechanisms on numerical data given in Section~\ref{sc:bg:dp}, some information on the data is required, such as the diameter 
$diam$ for the Laplace mechanism, and the binary values for the binary mechanism.
This information must somehow be provided to the \CSP for the mechanisms to be implemented.
As we will show later, it is sufficient for this information to be available in encrypted form.
Making such information about the data publicly available may reveal unwanted information and lead to inference attacks (e.g.\ attribute type, extreme values, etc.), and is therefore not desirable.

Before the data is encrypted, the \DO selects lower and upper bounds $l \le u \in \mathbb{R}$ that are independent of the data.
This may be performed by examination of the attribute in question (e.g.\ a person's age), or by other means, but must not be a function of the data (i.e.\ the range of the data).
In the case of binary-valued data, $l$ and $u$ will simply be the two binary values.
Non-informative bounding, as discussed in~\cite{Liu17}, ensures no additional privacy leakage, allowing the entire privacy budget $\epsilon$ to be spent on the differential privacy mechanism itself.
These bounds must then be encrypted and stored securely alongside the dataset in question.
For the remainder of this subsection, we will refer to the encrypted values $l^* = Enc_{pk}(l)$ and $u^* = Enc_{pk}(u)$.



Below, we detail how we can use $FHE.Add$ and $FHE.Mult$ (Section~\ref{subsec:swhe}) to render the encrypted values differentially private, without having to decrypt the original values.
This process is then applied independently to each value of interest.

\begin{itemize}

\item {\bf Laplace mechanism}: 
To achieve differential privacy, the required scale factor is $b = \frac{diam}{\epsilon}$.
In determining the noise to add to the data, we sample $L \sim \Laplace(0, b)$, and add this to the encrypted value.
In generating $L$, we draw a value $r$ at random from a uniform distribution on $\left[-\frac{1}{2}, \frac{1}{2}\right]$, $r \sim \Uniform\left(-\frac{1}{2}, \frac{1}{2}\right)$, and use the inverse of the cumulative probability distribution of $\Laplace(0, b)$ to find
$$L = - b \sgn(r) \log(1-2|r|),$$
where $\sgn(\cdot)$ is the signum function, defined by
$$\sgn(r) = \begin{cases} 1, & r > 0,\\ 0, & r =0,\\ -1, & r<0.\end{cases}$$

We cannot calculate the plaintext $L$, since $diam$ can only be calculated in encrypted form.
We can, however, calculate its ciphertext $L^* = Enc_{pk}(L)$ as :
\begin{multline*}
\hspace*{-10pt} L^* = 
Mult_{pk}\left(diam^*, Enc_{pk}\left(-\frac{1}{\epsilon} \sgn(r) \log(1-2|r|)\right)\right),$$
\end{multline*}
where $diam^* = Enc_{pk}(diam)$ is given by:
$$diam^* = Add_{pk}(u^*, Mult_{pk}(Enc_{pk}(-1), l^*)).$$

The resultant value that is stored is therefore
$$Add_{pk}(d^*_i, L^*).$$

%
%

\item {\bf Binary mechanism}:
If the original data $d$ is binary, the binary mechanism can be used.
This time we draw $r$ at random from the unit interval $[0,1]$.
If $r \le \frac{e^\epsilon}{1 + e^\epsilon}$, then the value $d_i^*$ remains unchanged.
However, if $r > \frac{e^\epsilon}{1 + e^\epsilon}$, then we flip $d_i^*$ by setting $d_i^\prime = u + l - d_i^*$.
Again, this can be done without knowing the value of $d_i^*$, and by only knowing the ciphertexts $l^*$ and $u^*$.
We can implement this using $FHE.Mult$ to get $-d^*_i$, and then using $FHE.Add$ as before.

In the case of the value being flipped, the value that is stored is
$$Add_{pk}(Mult_{pk}(d^*_i, Enc_{pk}(-1)), Add_{pk}(l^*, u^*)).$$


\end{itemize}

\subsection{k-Anonymization}
\label{subsec:anon}

In this paper, we implement anonymization algorithms that support the \emph{$k$-anonymity} privacy guarantee as formally defined in \cite{sweeneykanon}.
Given the identification threshold $k$, achieving \emph{$k$-anonymity} over encrypted data is a three-step process. First, we securely partition the data into clusters. In this paper, we specifically apply $k$-means clustering algorithm over encrypted data and use \emph{Squared Euclidean Distance}  (SED) metric to calculate the proximity of values in their respective feature space.
Second, to ensure that each cluster has at least $k$ members we apply data suppression and re-assignment techniques as presented in Sections~\ref{subsubsec:clustersupp} and \ref{subsubsec:clusterreassign}, respectively.  And finally, we securely anonymize the original data values to a representative one. For the numerical attributes, we replace them with the cluster centroid.  For each categorical attribute, we replace them with the common ancestor of the attribute value based on the respective generalization hierarchy.

\begin{algorithm}[h!]
\caption{\bf \textit{Secure k-Anonymization}}
\label{alg:kanon}
\small
\begin{algorithmic}[1]
  
\REQUIRE \CA has $T^{*}$, $k$, \emph{rounds} and \emph{th}
\ENSURE \CA computes the k-anonymized dataset $T^{*'}$

\STATE $k' \leftarrow N/k$

\STATE $\mathcal{C}^{*} \leftarrow \{c^{*}_1, \cdots, c^{*}_{k'}\}$, $c^{*}_i \in_R T^{*}$
\label{alg:kanon:icc}

\STATE \emph{loop} $\leftarrow 0$

\WHILE {\emph{loop} $<$ \emph{rounds}}

\label{alg:kanon:sed_start}
\FOR{$i=1$ to $N$}
\FOR{$j=1$ to $k'$}
\STATE $\mathcal{D}^{*}_{ij} \leftarrow \sum_{l = 1}^{d} (\mathcal{T}^{*}_{il} - \mathcal{C}^{*}_{jl})^2 $
\label{alg:kanon:sed}
\ENDFOR
\ENDFOR
\label{alg:kanon:sed_end}

\FOR{$i=1$ to $N$}
\STATE $\mathcal{I}^{*}_i \leftarrow $ {\bf \emph{ComputeMinIndex}}($\mathcal{D}^{*}_{i}$) 
\label{alg:kanon:new_cluster}
\ENDFOR

\STATE $\mathcal{C}^{*} \leftarrow $ {\bf \emph{RecomputeClusterCentres}}($\mathcal{I^{*}}$)
\label{alg:kanon:cluster_rec}

\STATE \emph{loop} $ = $ \emph{loop} $ + 1$ 

\ENDWHILE 

\STATE $\mathcal{J} \leftarrow $ {\bf \emph{Non-kClusters} ($\mathcal{I^{*}}$)}
\label{alg:kanon:nonk}

\STATE $\mathcal{I}^{*}, \mathcal{J} \leftarrow $ {\bf \emph{SupressClusters}}($\mathcal{I}^{*}$, $\mathcal{J}$, \emph{th}) 
\label{alg:kanon:supp}


\STATE $\mathcal{I}^{*} \leftarrow $ {\bf \emph{ReAssignClusters}}($\mathcal{I}^{*}$, $\mathcal{J}$)
\label{alg:kanon:reassign}

\STATE $T^{*'}$ $\leftarrow$ {\bf \emph{AnonymizeClusters}}($T^{*}$, $\mathcal{I}^{*}$)
\label{alg:kanon:anon}
\end{algorithmic}
\end{algorithm}

Algorithm~\ref{alg:kanon} outlines the procedure to compute $k$-anonymized data for a database table $\mathcal{T}$ having $d$ attributes. 
The algorithm takes as input the encrypted table $T^* = Enc_{pk}(T)$, the identification threshold $k$, the number of iterations of clustering algorithm \emph{rounds} and the suppression threshold \emph{th} at \CA. Further \CB has the secret key $sk$. In the end, the algorithm outputs the corresponding $k$-anonymized database table $T^{*'}$.

In the following sections, we will describe different steps of Algorithm~\ref{alg:kanon}.

\subsubsection{Data Clustering}
\label{subsubsec: dataclustering}
To produce a $k$-anonymized database, \CA can at most find $k' = N/k$ clusters, each having at-least $k$ members, where $N$ is the total number of tuples in the table $\mathcal{T}^*$. In Step~\ref{alg:kanon:icc} we randomly select $k'$ tuples as the initial cluster centres\footnote{Other initial cluster center selection methods can be used} and in Steps~\ref{alg:kanon:sed_start} --~\ref{alg:kanon:sed_end}, squared Euclidean distance is computed for all the tuples from all the cluster centres using the homomorphic properties of the \SWHE encryption scheme and the results are stored in matrix $\mathcal{D}^*$. 

Next, in Step~\ref{alg:kanon:new_cluster} new cluster assignment for all the tuples are identified by calling the function {\bf \emph{ComputeMinIndex}} described in Algorithm~\ref{alg:minIndex}. 
This algorithm takes as input a vector $\mathcal{D}^*_i$ of size $k'$ and returns an encrypted vector of size $k'$ having the encryption of value $1$ at the index of nearest cluster centre and encryption of $0$ at all other positions. In Steps~\ref{alg:minIndex:poly_start} --~\ref{alg:minIndex:poly_end}, \CA selects a monotonic increasing polynomial $poly(x)$, such that $poly(x_1) \geq poly(x_2) $ iff $x_1 \geq x_2$ and homomorphically evaluate the polynomial $poly(x)$ over all the values in vector $\mathcal{D}^*_i$ and computes $\mathcal{D}^{*'}_i$. Next, in Step~\ref{alg:minIndex:perm} \CA selects a pseudo-random permutation (PRP) $\pi$ and permutes the vector $\mathcal{D}^{*'}_i$.  Finally, it sends the vector $\mathcal{D}^{*''}_i$ to \CB.
Then, in Step~\ref{alg:minIndex:dec}, \CB decrypts the vector $\mathcal{D}^{*''}_i$ and in Steps~\ref{alg:minIndex:min_start} --~\ref{alg:minIndex:min_end} identifies the index of the minimum value element in vector $\mathcal{D}^{''}_i$. In Steps~\ref{alg:minIndex:ret_start} --~\ref{alg:minIndex:ret_end}, \CB initializes a vector $\mathcal{I}^{'}_i$ of size $k'$ with $0$ and then sets the value at the index identified above to $1$. Then, in Step~\ref{alg:minIndex:ret_enc} \CB encrypts the vector $\mathcal{I}^{*'}_i$ and sends it to \CA. Note that the vector $\mathcal{I}^{*'}_i$ contains the encryption of $1$ at exactly one position corresponding to the nearest cluster center, but since the vector $\mathcal{D}^{*''}_i$ was initially permuted by \CA, hence \CB does not learns the correct cluster assignment.
Next, \CA applies the inverse permutation $\pi^{-1}$ to $\mathcal{I}^{*'}_i$ in Step~\ref{alg:minIndex:inv_perm}. Note, \CA has received the cluster assignment for tuple $T^{*}_i$ but it does not learn the cluster to which this tuple is assigned, since all the entries in the vector $\mathcal{I}^{*}_i$ are encrypted using non-deterministic encryption.
Similarly, \CA receives the encrypted cluster assignment for every tuple in the encrypted table $\mathcal{T}^*$

\begin{algorithm}[h!]
\small
\caption{\bf \textit{ComputeMinIndex}}
\label{alg:minIndex}
\begin{algorithmic}[1]
\setlength{\itemsep}{-1pt}
\REQUIRE \CA has $\mathcal{D}^{*}_{i}$
\ENSURE \CA gets an encrypted vector $\mathcal{I}^{*}_i$,  having value $1$ at the index of nearest cluster center and $0$ otherwise

\STATE \CA:
\label{mi:a1}
\vspace*{-4pt}

\begin{enumerate}
\setlength{\itemsep}{-1pt}

\item Choose a polynomial $poly(x) \leftarrow a_0 + a_1 \cdot x + \cdots + a_q \cdot x^q$, $q,a_l \in_R \mathbb{N} ~\forall l \in \{0,q\}$ 
\label{alg:minIndex:poly_start}

\item  $\mathcal{D}^{*'}_{ij} \leftarrow poly(\mathcal{D}^{*}_{ij}) ~\forall j \in \{1,k'\}$
\label{alg:minIndex:poly_end}

\item $\mathcal{D}^{*''}_{i} \leftarrow \pi_{s}(\mathcal{D}^{*'}_{i})$, $\pi : \{0,1\}^{k'} \times \{0,1\}^{s} \rightarrow \{0,1\}^{k'}$
\label{alg:minIndex:perm}

\item {\bf Send} $\mathcal{D}^{*''}_{i}$ to \CB

\end{enumerate}
\vspace*{-4pt}

\STATE \CB:
\label{mi:b1}
\vspace*{-4pt}

\begin{enumerate}
\setcounter{enumi}{4}
\setlength{\itemsep}{-1pt}

\item $\mathcal{D}^{''}_{ij} \leftarrow Dec_{sk}(\mathcal{D}^{*''}_{ij}) ~\forall j \in \{1,k'\}$
\label{alg:minIndex:dec}

\item $ min\_ind \leftarrow 1$, $min\_val \leftarrow \mathcal{D}^{''}_{i1}$
\label{alg:minIndex:min_start}

\item {\bf for} $ j = 2 $ to $k'$  {\bf do}

    \item \hspace*{10pt} {\bf if} $\mathcal{D}^{''}_{ij} < min\_val$ {\bf then}
        
        \item \hspace*{20pt} $min\_val \leftarrow  \mathcal{D}^{''}_{ij}$        
        
        \item \hspace*{20pt} $min\_ind \leftarrow j$
    
    \item \hspace*{10pt} {\bf end if}    
    
\item {\bf end for}
\label{alg:minIndex:min_end}

\item $\mathcal{I}^{'}_i \leftarrow \vec{0}_{k'} $
\label{alg:minIndex:ret_start}

\item $\mathcal{I}^{'}_{i}[min\_ind] \leftarrow 1$
\label{alg:minIndex:ret_end}

\item $\mathcal{I}^{j*'}_{i} \leftarrow Enc_{pk}(\mathcal{I}^{j'}_{i}) ~\forall j \in \{1,k'\}$
\label{alg:minIndex:ret_enc}

\item {\bf Return} $\mathcal{I}^{*'}_i$ to \CA
\label{alg:minIndex:ret}

\end{enumerate}
\vspace*{-4pt}

\STATE \CA:
\label{mi:a2}
\vspace*{-4pt}

\begin{enumerate}
\setcounter{enumi}{16}
\setlength{\itemsep}{-1pt}

\item $\mathcal{I}^{*}_i \leftarrow \pi^{-1}_{s}(\mathcal{I}^{*'}_{i})$
\label{alg:minIndex:inv_perm}

\end{enumerate}

\end{algorithmic}
\end{algorithm}

Next, in Step~\ref{alg:kanon:cluster_rec} of Algorithm~\ref{alg:kanon} \CA calls the function {\bf \emph{RecomputeClusterCentres}} to recompute the cluster center representatives. Algorithm~\ref{alg:newclustercenter} provides the details of this function. It takes as input the encrypted cluster assignment $\mathcal{I^{*}}$ and returns new cluster centres $\mathcal{C}^{*}$. In Steps~\ref{alg:newclustercenter:count_start} --~\ref{alg:newclustercenter:count_end}, Algorithm~\ref{alg:newclustercenter} first computes the encrypted cluster count and sum for every cluster. Note in Step~\ref{alg:newclustercenter:sum}, tuple $i$ is added to $sum^*_j$ if and only if it belongs to a cluster $j$, since  $\mathcal{I^{*}}_{ij} = Enc_{pk}(1)$ if tuple $i$ belongs to a cluster $j$, else $\mathcal{I^{*}}_{ij} = Enc_{pk}(0)$. Next, in Step~\ref{alg:newclustercenter:count_blind}, \CA selects a  random value $u_j$ and multiplies it with the cluster count for cluster $j$. This step produces a one time pad blinding of the cluster count. Similarly in Step~\ref{alg:newclustercenter:sum_blind} the corresponding cluster sum is blinded with a random value $v_j$. All the above operations are performed using the homomorphic properties of the \SWHE encryption scheme. Now, \CA sends the blinded cluster count and sum to \CB. 
In Steps~\ref{alg:newclustercenter:div_start} --~\ref{alg:newclustercenter:div_end} \CB decrypts and divides the corresponding cluster sum and count and re-encrypts the results. \CB then sends the encrypted divisions to \CA.
In Steps~\ref{alg:newclustercenter:rem_blind}, \CA multiplies the cluster divisions with 
division of the random values selected in Steps~\ref{alg:newclustercenter:count_blind} and \ref{alg:newclustercenter:sum_blind}. This step removes the randomness and \CA gets the updated cluster centres encrypted under the public key $pk$.
The clustering process is repeated for \emph{rounds} number of iterations in order to converge the cluster centres.

\begin{algorithm}[h!]
\small
\caption{\bf \textit{RecomputeClusterCentres}}
\label{alg:newclustercenter}
\begin{algorithmic}[1]

\REQUIRE Cluster assignment $\mathcal{I^{*}}$
\ENSURE  Returns new cluster centres $\mathcal{C}^{*}$
\setlength{\itemsep}{-1pt}

\STATE \CA:
\label{rcc:a1}
\vspace*{-4pt}

\begin{enumerate} 
\setlength{\itemsep}{-1pt}

\item {\bf for} $j = 1$ to $k'$ {\bf do}

    \item \hspace*{10pt}  ${count}^{*}_j \leftarrow  Enc_{pk}(0) $; ${sum}^{*}_j \leftarrow  Enc_{pk}(0)$ 

    \item \hspace*{10pt} {\bf for} $i = 1$ to $N$ {\bf do}
    \label{alg:newclustercenter:count_start}
    
        \item \hspace*{20pt} ${count}^{*}_j $ += $\mathcal{I}^{*}_{ij}$
        
        \item \hspace*{20pt} ${sum}^{*}_j $ += $\mathcal{I}^{*}_{ij} * \mathcal{T}^{*}_{i}$
        \label{alg:newclustercenter:sum}

    \item \hspace*{10pt} {\bf end for}
    \label{alg:newclustercenter:count_end}
        
    \item \hspace*{10pt} ${count}^{*'}_j \leftarrow {count}^{*}_j $ * $Enc_{pk}(u_j)$; $u_j \in_R \mathbb{N}$
    \label{alg:newclustercenter:count_blind}

    \item     \hspace*{10pt} ${sum}^{*'}_j \leftarrow {sum}^{*}_j $ * $Enc_{pk}(v_j)$; $v_j \in_R \mathbb{N}$
    \label{alg:newclustercenter:sum_blind}
        
\item {\bf end for}

\item Choose a PRP $\pi_{s} : \{0,1\}^{k'} \times \{0,1\}^{s} \rightarrow \{0,1\}^{k'}$
\label{alg:newclustercenter:perm}

\item ${sum}^{*''} \leftarrow \pi_{s}({sum}^{*'})$;  ${count}^{*''} \leftarrow \pi_{s}({count}^{*'})$
\label{alg:newclustercenter:apply_perm}

\item {\bf Send } ${sum}^{*''}$ and ${count}^{*''}$ to \CB

\end{enumerate}

\vspace*{-4pt}
\STATE \CB:
\label{rcc:b1}
\vspace*{-4pt}\\

\begin{enumerate}
\setcounter{enumi}{12}
\setlength{\itemsep}{-1pt}

\item {\bf for} $j = 1$ to $k'$ {\bf do}
\label{alg:newclustercenter:div_start}

    \item \hspace*{10pt} $div^{*''}_j \leftarrow Enc_{pk}(Dec_{sk}(sum^{*''}_j) / Dec_{sk}(count^{*''}_j))$

\item {\bf end for}
\label{alg:newclustercenter:div_end}

\item {\bf Return} $div^{*''}$ to \CA

\end{enumerate}

\vspace*{-4pt}
\STATE \CA:
\label{rcc:a2}
\vspace*{-4pt}

\begin{enumerate}
\setcounter{enumi}{16}
\setlength{\itemsep}{-1pt}

\item ${div}^{*'} \leftarrow \pi^{-1}_{s}({div}^{*''})$
\label{alg:newclustercenter:rem_perm}

\item {\bf for} $j = 1$ to $k'$ {\bf do}
    
    \item \hspace*{10pt} $\mathcal{C}^{*}_j \leftarrow Enc_{pk}(u_j / v_j) * div^{*'}_j$
    \label{alg:newclustercenter:rem_blind}

\item {\bf end for}

\end{enumerate}

\end{algorithmic}
\end{algorithm}

\begin{figure*}[!ht]
    \centering
     \subfloat[Cluster to Cluster\label{subfig:s1}]{%
      \includegraphics[width=.20\linewidth, height=3.5cm]{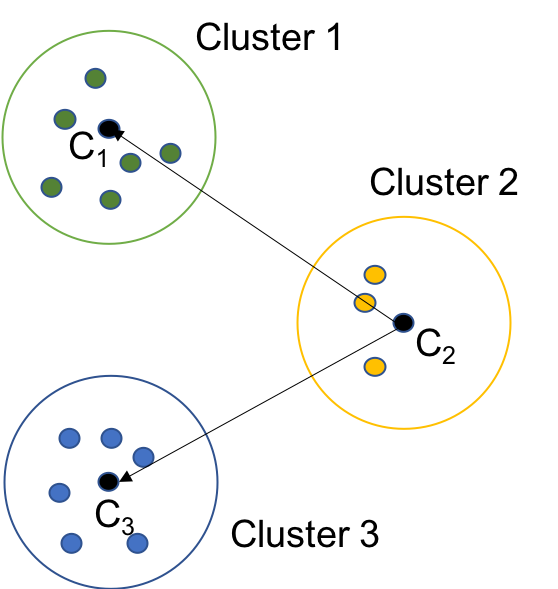}
     }
     \subfloat[Point to Cluster\label{subfig:s2}]{%
      \includegraphics[width=.20\linewidth, height=3.5cm]{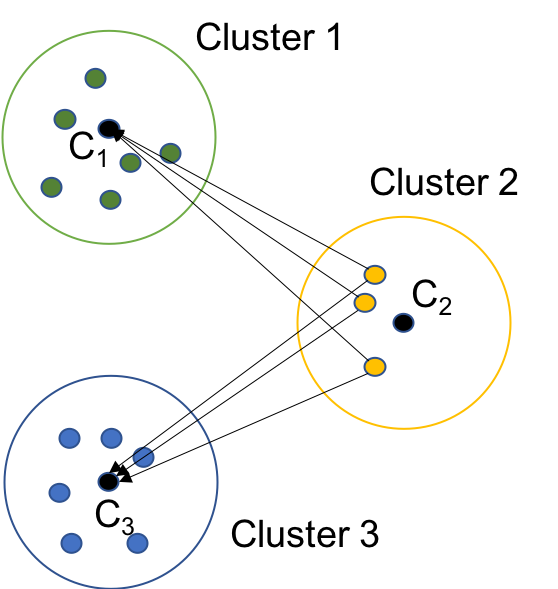}
     }
     \subfloat[Point to Point\label{subfig:s3}]{%
      \includegraphics[width=.20\linewidth, height=3.5cm]{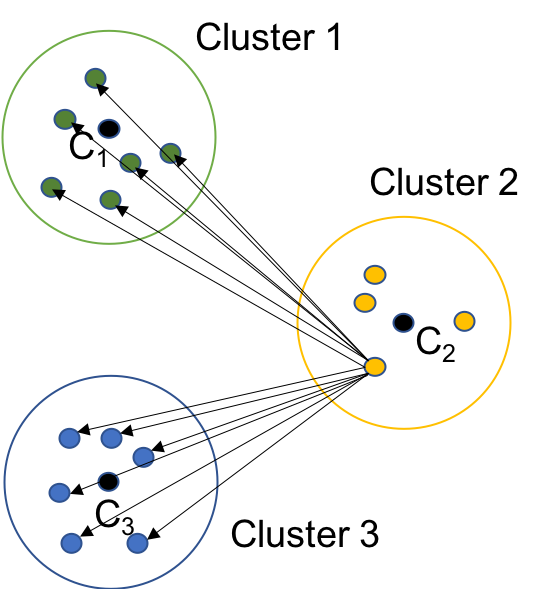}
     }
     \caption{Cluster Re-assignment Strategies}
     \label{fig:reassgn}
  \end{figure*}

\subsubsection{Cluster Suppression}
\label{subsubsec:clustersupp}
The above clustering process does not provide guarantee on the number of members in each cluster.  In order to achieve \emph{$k$-anonymity}, we make sure that each cluster has at-least $k$ members. To achieve this, we apply a post-processing phase on the output clusters. In Step~\ref{alg:kanon:nonk}, Algorithm~\ref{alg:kanon} calls function {\bf \emph{Non-kClusters}} to identify the clusters having fewer than $k$ members. The details of this function is described in Algorithm~\ref{alg:nonk}. It takes as input the encrypted cluster assignment ($\mathcal{I^{*}}$) and encrypted cluster count (${count}^{*}$) and returns a vector $\mathcal{J}$ of size $k'$ indicating the clusters which need further processing. In Step~\ref{alg:nonk:poly} \CA selects a monotonically  increasing polynomial $poly(x)$ and homomorphically evaluate the polynomial over the encrypted cluster counts and computes ${count}^{*'}$. Next, in Step~\ref{alg:nonk:apply_perm} a PRP $\pi_{s}$ is selected and used to permute the order of ${count}^{*'}$. Then in Step~\ref{alg:nonk:enc_k}, the identification factor $k$ is encrypted and masked with the polynomial $poly(x)$. Next both ${count}^{*'}$ and ${mark}^{*'}$ are sent to \CB.
Party \CB initializes a vector $\mathcal{J}^{'}$ of size $k'$ with value $0$. Then in Steps~\ref{alg:nonk:dec_k} --~\ref{alg:nonk:dec_count_end}, it sets the entry of vector $\mathcal{J}^{'}$ to $1$ if the count is less than $k$ and finally returns $\mathcal{J}^{'}$. Party \CA then applies the inverse permutation $\pi^{-1}_{s}$ and retrieves the vector $\mathcal{J}$.

\begin{algorithm}[h!]
\caption{\bf \textit{Non-kClusters}}
\label{alg:nonk}
\small
\begin{algorithmic}[1]

\REQUIRE Cluster assignment $\mathcal{I^{*}}$, Cluster counts ${count}^{*}$ 
\ENSURE \CA learns Non-k Clusters $\mathcal{J}$
\setlength{\itemsep}{-1pt}

\STATE \CA:
\vspace*{-4pt}

\begin{enumerate}
\setlength{\itemsep}{-1pt}

\item Choose a polynomial $poly(x) \leftarrow a_0 + a_1 \cdot x + \cdots + a_q \cdot x^q$, $q,a_l \in_R \mathbb{N} ~\forall l \in \{0,q\}$ 
\label{alg:nonk:poly}


%
%
%
        
\item ${count}^{*'}_j \leftarrow poly({count}^{*}_j)$, $\forall j \in [1,k']$
\label{alg:nonk:count_blind}



\item ${count}^{*''} \leftarrow \pi_{s}({count}^{*'})$ \COMMENT{$\pi_{s}$ is a PRP}
\label{alg:nonk:apply_perm}

\item ${mark}^* \leftarrow Enc_{pk}(k)$;  ${mark}^{*'} \leftarrow poly({mark}^*)$
\label{alg:nonk:enc_k}

\item {\bf Send } ${mark}^{*'}$ and ${count}^{*''}$ to \CB

\end{enumerate}

\vspace*{-4pt}
\STATE \CB:
\vspace*{-4pt}

\begin{enumerate}
\setcounter{enumi}{5}
\setlength{\itemsep}{-1pt}

\item Initialize vector $\mathcal{J}^{'} \leftarrow \vec{0}_{k'} $
\label{alg:nonk:init_j}

\item ${mark}^{'} \leftarrow Dec_{sk}({mark}^{*'})$
\label{alg:nonk:dec_k}

\item {\bf for} $j = 1$ to $k'$ {\bf do}
\label{alg:nonk:dec_count_start}

    \item \hspace*{10pt} ${count}^{''}_{j} \leftarrow Dec_{sk}({count}^{*''}_{j})$
    
    \item \hspace*{10pt} {\bf if} ${count}^{''}_{j} < {mark}^{'}$ {\bf then}
        
        \item \hspace*{20pt} $\mathcal{J}^{'}_{j} \leftarrow 1$
    
    \item \hspace*{10pt} {\bf end if}

\item {\bf end for}
\label{alg:nonk:dec_count_end}

\item {\bf Return} $\mathcal{J}^{'}$ to \CA

\end{enumerate}

\vspace*{-4pt}
\STATE \CA:
\vspace*{-4pt}

\begin{enumerate}
\setcounter{enumi}{14}
\setlength{\itemsep}{-1pt}

\item $\mathcal{J} \leftarrow \pi^{-1}_{s}(\mathcal{J}^{'} )$
\label{alg:nonk:result}

\end{enumerate}

\end{algorithmic}
\end{algorithm}

Now, if the number of members is more than $k$ (i.e.\ if $\mathcal{J} == 0$), then the cluster is left unmodified. If, however, the cluster contains fewer than $k$ members we follow two strategies. 
First, we check if we can suppress the cluster and remove its points from the final anonymized output. For suppression, a threshold is required to specify the maximum percentage of the total points we are allowed to remove. As an example, if the suppression threshold is 10\% and the input dataset contains 200 tuples, we are only allowed to remove up to 20 tuples. If suppression is not allowed or we have reached the suppression threshold, then we apply cluster re-assignment techniques, in which nearest clusters are identified to merge with the non-$k$ clusters.
The cluster re-assignment strategies are described in Section~\ref{subsubsec:clusterreassign}.

\subsubsection{Cluster Re-assignment}
\label{subsubsec:clusterreassign}
Once the suppression threshold is reached, the remaining non-$k$ clusters are re-assigned to nearest clusters. The merging of two clusters is easily done using the encrypted cluster assignment vector. For example, say we want to merge cluster $j$ in cluster $i$, we can achieve this by adding the $j$th component of every encrypted cluster assignment vector $\mathcal{I}^*$ to its $i$th component. 
Further, merging the cluster with its nearest one does not guarantee $k$-anonymity. For example, let's assume we want to achieve $3$-anonymity and a cluster has one member. Its nearest cluster also has a single member so merging them will not result in a cluster with a minimum of $3$ elements.  Thus, we need to apply the process iteratively until all created clusters have more than or equal to $k$ data points.
The nearest cluster could be identified using one of the following strategies:
\begin{enumerate}

\item {\bf Cluster to Cluster -- } The nearest cluster is computed based on the squared Euclidean distance of the target non-$k$ cluster centroid from the centroid of the rest of the clusters, as shown in Figure \ref{subfig:s1}.

\item {\bf Point to Cluster -- } The nearest cluster is computed based on the squared Euclidean distance of the data points in the target non-$k$ cluster from the centroid of the rest of the clusters, as shown in Figure \ref{subfig:s2}.

\item {\bf  Point to Point -- } The nearest cluster is computed based on the squared Euclidean distance of the data points in the target non-$k$ cluster from the data points in the rest of the clusters, as shown in Figure \ref{subfig:s3}.

\end{enumerate}

After cluster re-assignment step all the identified clusters have a minimum of $k$ data points.

\subsubsection{Data Anonymization}
\label{subsubsec:dataanon}

For numerical attributes, we replace them with the cluster centroid. 
For the categorical attributes, we replace them with the common ancestor of the attribute value based on the respective generalization hierarchy.
In order to avoid inference attacks based on the hierarchy structure and cardinality of number of nodes per level, we can employ similar approaches like with data masking dictionaries; we can randomly insert dummy nodes at each level.
One approach to calculate the common ancestor for all values is the following. We first calculate the common ancestor between the first value and the second one, let's call it
$A_{1,2}$. Then we calculate the common ancestor between $A_{1,2}$ and the third value, $A_{1,2,3}$ and so forth. If at some point one of the common ancestors calculated is the root of the hierarchy, then the calculations stop. This approach requires at maximum O(N) checks and each check requires $O(M)$ operations, where $N$ is the total number of values and $M$ is the height of the generalization hierarchy. 

We will now describe how we calculate the common ancestor between two values. We will use the hierarchy of Figure \ref{fig:cat} as an example. Let us consider the root of the hierarchy to be level 2.
We begin from the fact that all the encrypted values in the data will belong to the leaves of the hierarchy (level 0). 
Given two encrypted values from the data, {\it v1} and {\it v2}, we find the nearest node from level 1 for each value using a secure kNN approach\cite{secureKNN, manish} with $k=1$. Let us call the nearest nodes
$N_1(v1)$ and $N_1(v2)$. We subtract the values $N_1(v1)$ and $N_1(v2)$ and we forward the difference to Party P2. If the distance is zero, then it means it is the same node and thus we 
found the common ancestor. If the difference is non-zero, we then follow the same process for $N_1(v1)$ and $N_1(v2)$ and we find their nearest nodes from the next level and so on. 
As an optimization, whenever we want to calculate the common ancestor of two values, we look immediately for the maximum level stopped at the previous steps. The entire process stops if for any
given pair we reach the root level.


\subsection{Risk and utility assessment}
\label{subsec:riskutility}

In this Section, we sketch out how various risk and utility assessment algorithms can be implemented on top of encrypted data. 

Inference-based risk metrics, such as the ones described in~\cite{chen1998estimation,hoshino2001applying,zayatz1991estimation} rely solely on the
size of the equivalence classes and additional external information, such as population size (required) and bias estimation (optional). Thus, it is only required to 
group the data based on their equivalence class and count the size of each group. 

Simple information loss metrics, such as Average Equivalence Class Size (AECS)~\cite{mondrian} and discernibility~\cite{discernibilitymetric},  also rely on the equivalence class size to provide a result. 
Categorical precision~\cite{precisionmetric} relies on the level of generalization applied for each value. This information is computed when we perform the data anonymization step (see Section~\ref{subsubsec:dataanon}). Similarly, generalized loss metric~\cite{glmmetric} requires the number of leaves for each anonymized value. 
However, metrics like non-uniform entropy~\cite{nonuniformentropymetric} and global certainty penalty~\cite{gcpmetric} require either frequency calculations 
or knowledge of the data diameter, which can be only aquired with access to the original values.


\section{Security Guarantees}
\label{sec:security}
As described earlier both the Parties~\CA and~\CB are considered in the \hbc security model where both the parties correctly execute the protocol but may try to learn the plaintext value from their view of the encrypted data processing. We also assume that Party~\CA and Party~\CB do not collude. Further, Party~\CB is additionally trusted with the secret key of the SHE encryption scheme. We want to emphasize that this cloud model is not new and has been used in related problem domain \cite{secureKNN, manish}.

Given above assumptions, informally we will prove that,
\emph{the views of Party~\CA and Party~\CB does not reveal any useful information about the plaintext database during the execution of secure $k$-Anonymization protocol}.
We will formally prove this statement using Leakage Profile Analysis.

\subsection{Leakage profile at Party \CA}
Below we enumerate the leakage to Party~\CA :

\begin{enumerate}

\item {\bf Direct Identifier : } In Algorithm~\ref{alg:direct}, for each encrypted value in the attribute, Party~\CA computes its difference from the remaining $N - 1$ values and then multiplies them with a different random value. Both these operations are performed using the homomorphic properties of the SHE scheme. Hence any leakage in this step will break the security guarantee of the underlying SHE encryption scheme.

\item {\bf ComputeMinIndex : } In Algorithm~\ref{alg:minIndex},  for each entry in vector $D_i^*$, Party~\CA evaluates a randomly chosen polynomial using the homomorphic properties of the SHE scheme. Now, since the polynomial evaluation is done over encrypted data, hence the security guarantee of the SHE scheme ensures that there is no leakage to Party~\CA. Next \CA chooses a pseudo-random permutation to hide the physical order of elements in vector  $D_i^{*'}$. This step further breaks any physical order co-relation between the entries in different $D_i^*$ vectors.

\item {\bf RecomputeClusterCentres : } In Algorithm~\ref{alg:newclustercenter}, Party~\CA computes the number of data points in every cluster and the corresponding cluster sum. To compute the count, \CA applies addition operation over the encrypted cluster assignment vector $\mathcal{I}^*$ and to compute the cluster sum, \CA first multiplies the encrypted vector $\mathcal{I}^*$ and encrypted data points $\mathcal{T}_i^{*}$ and then adds the encrypted values. All the operations in this algorithm are performed over encrypted data using the properties of the SHE scheme, hence the security guarantee of the SHE scheme ensures that there is no possible leakage to Party~\CA.

\item {\bf Non-kClusters : } In Algorithm \ref{alg:nonk}, Party \CA evaluates a randomly chosen polynomial over the encrypted cluster count values and the encrypted identification factor $k$, using the homomorphic properties of the SHE scheme. Hence, any leakage in these steps will break the security guarantee of the SHE scheme.

\item {\bf Suppress and Reassign Clusters : } In this step, Party \CA only replaces some encrypted cluster count values with random values or adds two encrypted vectors , hence there is no extra leakage.

\end{enumerate}

The above leakage profile for Party \CA leads to the following security guarantee :

\begin{Theorem}{\textnormal{\textbf{Security Guarantee for Party \CA:}}}
\emph{The secure $k$-Anonymization protocol leaks no information to Party \CA except that it learns if an attribute is a direct identifier and number points in the non-k clusters. In particular, Party~\CA does not gain any knowledge about the encrypted data points, the difference between two data points, the cluster to which a data point is assigned and the cluster centre representatives.}
\end{Theorem}

\subsection{Leakage profile at Party \CB}
Below we enumerate the leakage to Party~\CB :

\begin{enumerate}

\item {\bf Direct Identifier :} In Algorithm \ref{alg:direct}, Party \CB decrypts the encrypted matrix $\mathcal{M}$ using the secret key $sk$. But since  
Party~\CA has multiplied each entry of the matrix $\mathcal{M}$ with a different random value before sending it to Party \CB, the decrypted matrix $\mathcal{M}$ effectively contains random values. Hence the only leakage in this step is that Party \CB learns if two values in the attribute are equal since the corresponding decrypted difference will be $0$ but nothing is revealed about the original data points or the difference between two unequal values.

\item {\bf ComputeMinIndex : }
\label{sec:pb:CompluteMinIndex}
 In Algorithm \ref{alg:minIndex},  for every data point, Party \CB receives a vector $D_i^{*''}$ of size $k'$. The entries in this vector are the output of encrypted polynomial evaluation \emph{poly(x)} over the distance of the data point $i$ from the cluster centres. Further, the order of elements in the vector is permuted using a secure pseudo-random permutation, hence the exact identity of cluster centres associated with any given difference value is hidden from Party \CB. 

Party \CB decrypts the entries in the vector  $D_i^{*''}$ and since the polynomial \emph{poly(x)} is order preserving, hence Party \CB can sort the decrypted values and identify the index of the nearest cluster centre.   

In Appendix \ref{sec:ordeq}, we prove that recovering the plaintext distances form $D_i^{''}$ is computationally infeasible for Party \CB.
The only possible leakage to Party \CB in this round is the presence of such points in the database that are equidistant from two or more cluster centres. This is leaked from the presence of identical values in the set $\{poly(d^{''}_{i,1}), poly(d^{''}_{i,2}),\cdots, poly(d^{''}_{i,k'})\}$. However, since the order of the values is randomly permuted by Party \CA\!\!, Party \CB cannot map these values back to the original index of either the data point or the corresponding cluster centres in the database.

\begin{figure*}[tb!]
		\centering
		\subfloat[Data Encryption Time]{\label{fig:dp-encrypt}\includegraphics[width=.33\textwidth]{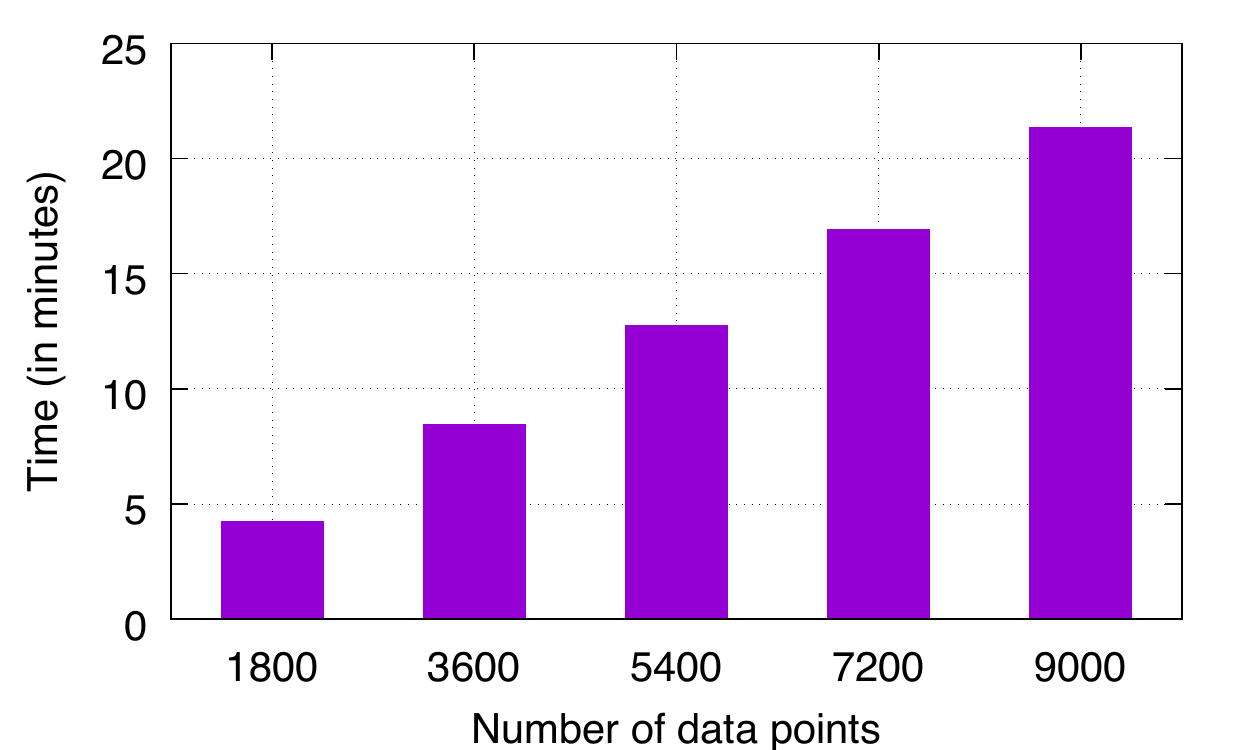}}
		\subfloat[Time to check an identifier]{\label{fig:dp-check-identifier}\includegraphics[width=.33\textwidth]{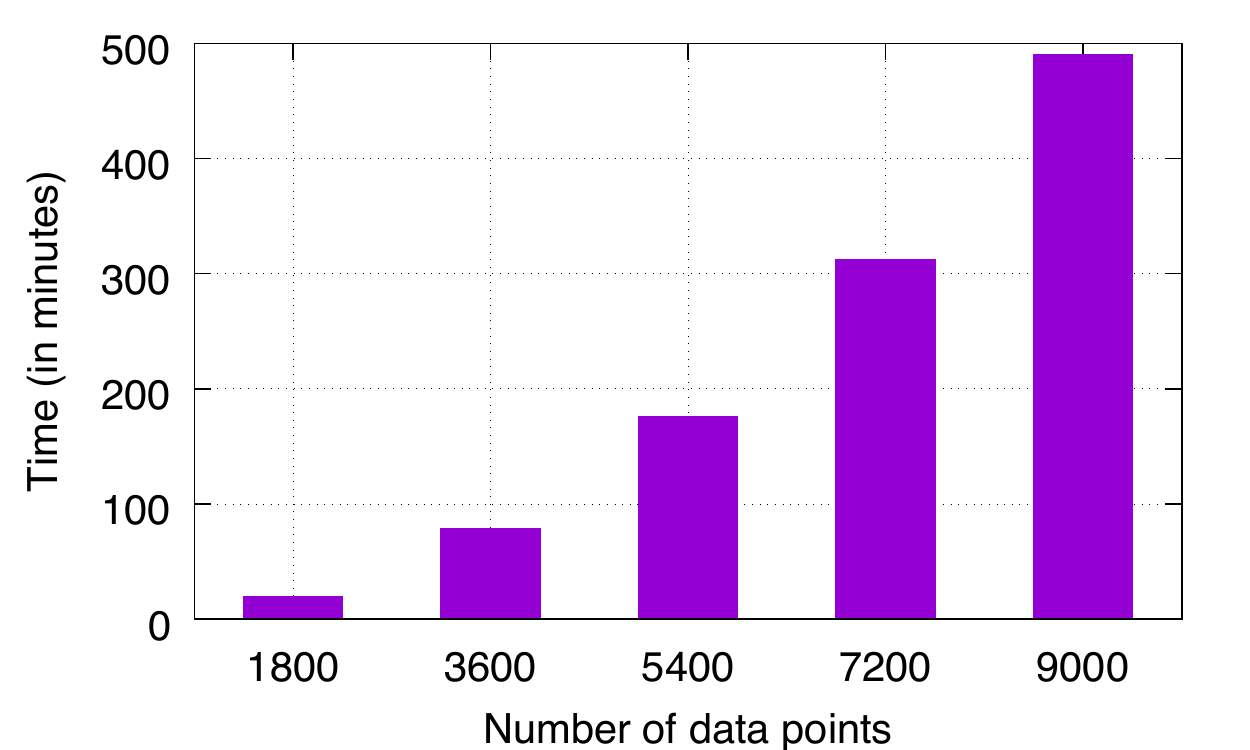}}
		\subfloat[Time taken per iteration of clustering]{\label{fig:dp-clustering-per-iteration}\includegraphics[width=.33\textwidth]{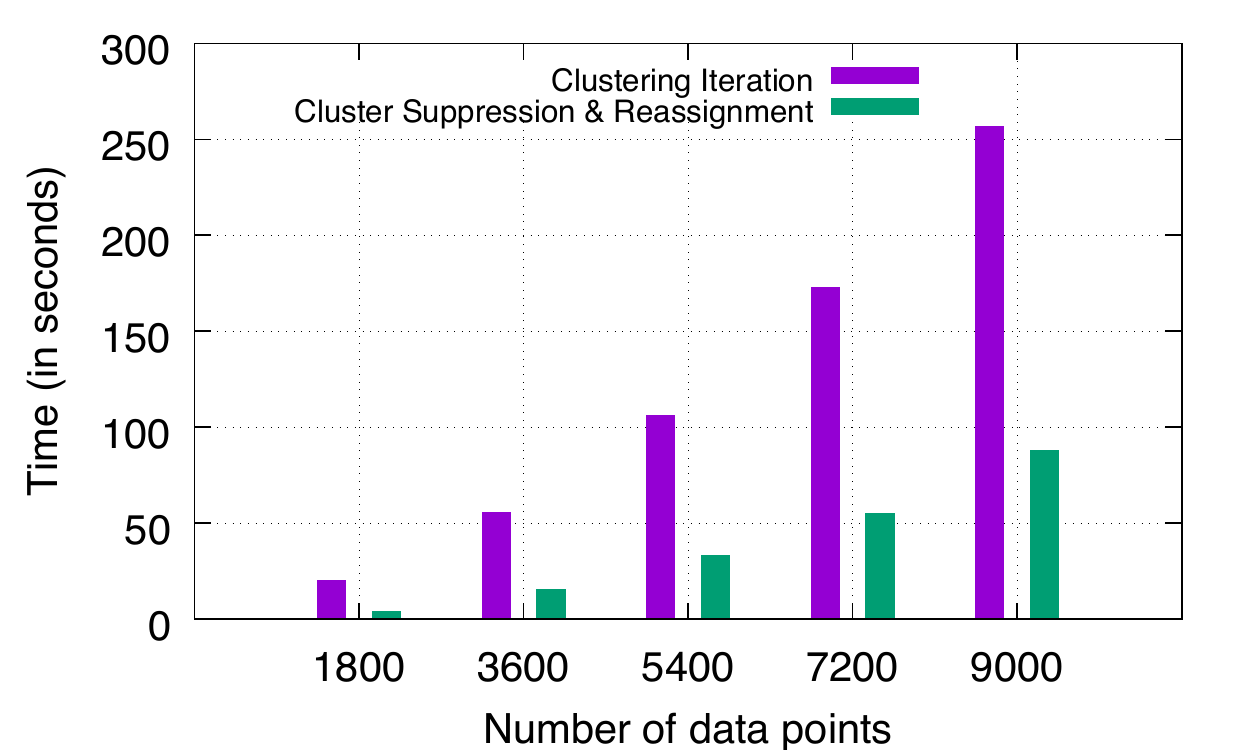}}

	\caption{Execution time for varying number of data points (each having 2 dimensions)}
	\label{fig:vary-data-points}
\end{figure*}

\begin{figure*}[tb!]
  \centering
		\subfloat[Data Encryption Time]{\label{fig:dim-encrypt}\includegraphics[width=.33\textwidth]{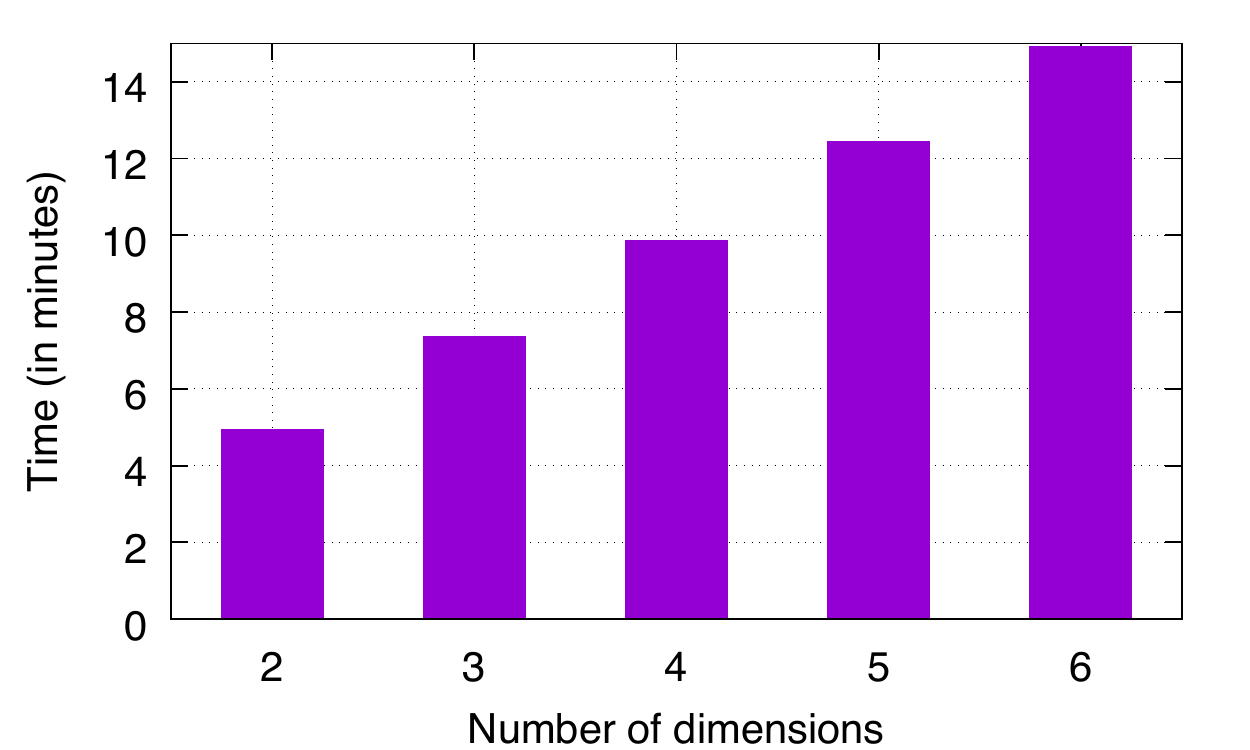}}
		\subfloat[Time to check an identifier]{\label{fig:dim-check-identifier}\includegraphics[width=.33\textwidth]{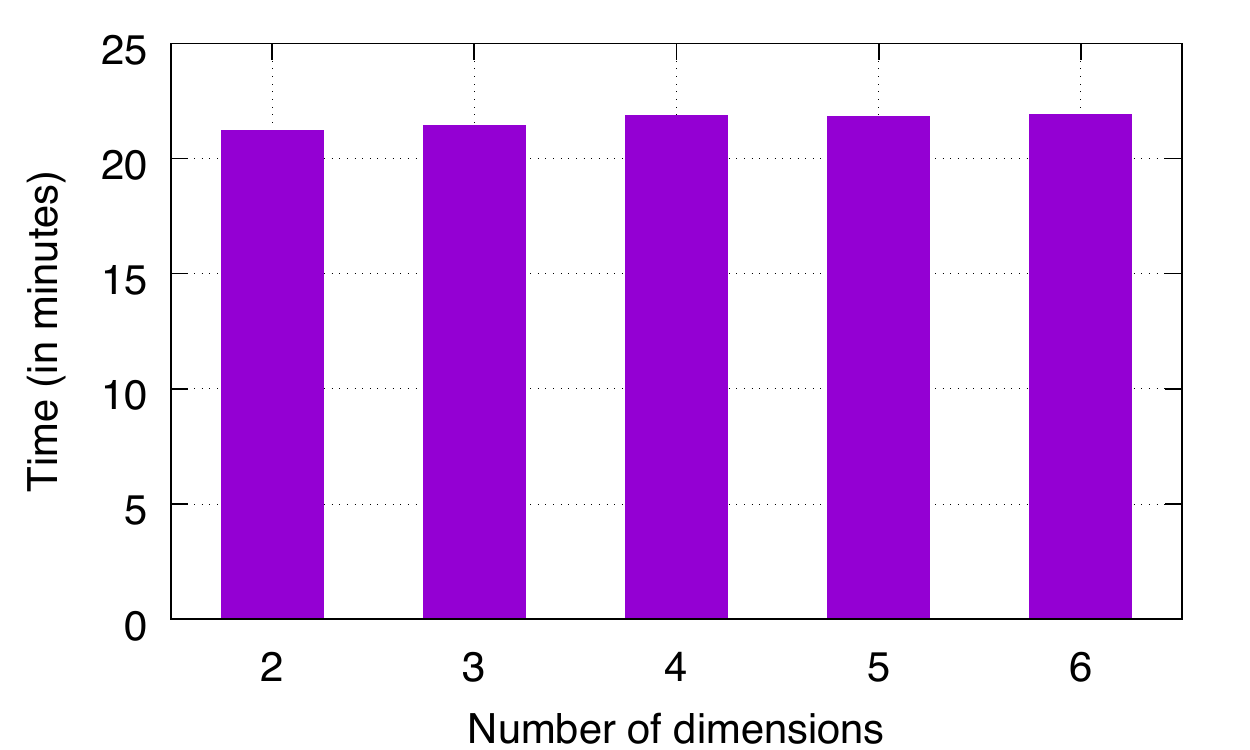}}
		\subfloat[Time taken per iteration of clustering]{\label{fig:dim-clustering-per-iteration}\includegraphics[width=.33\textwidth]{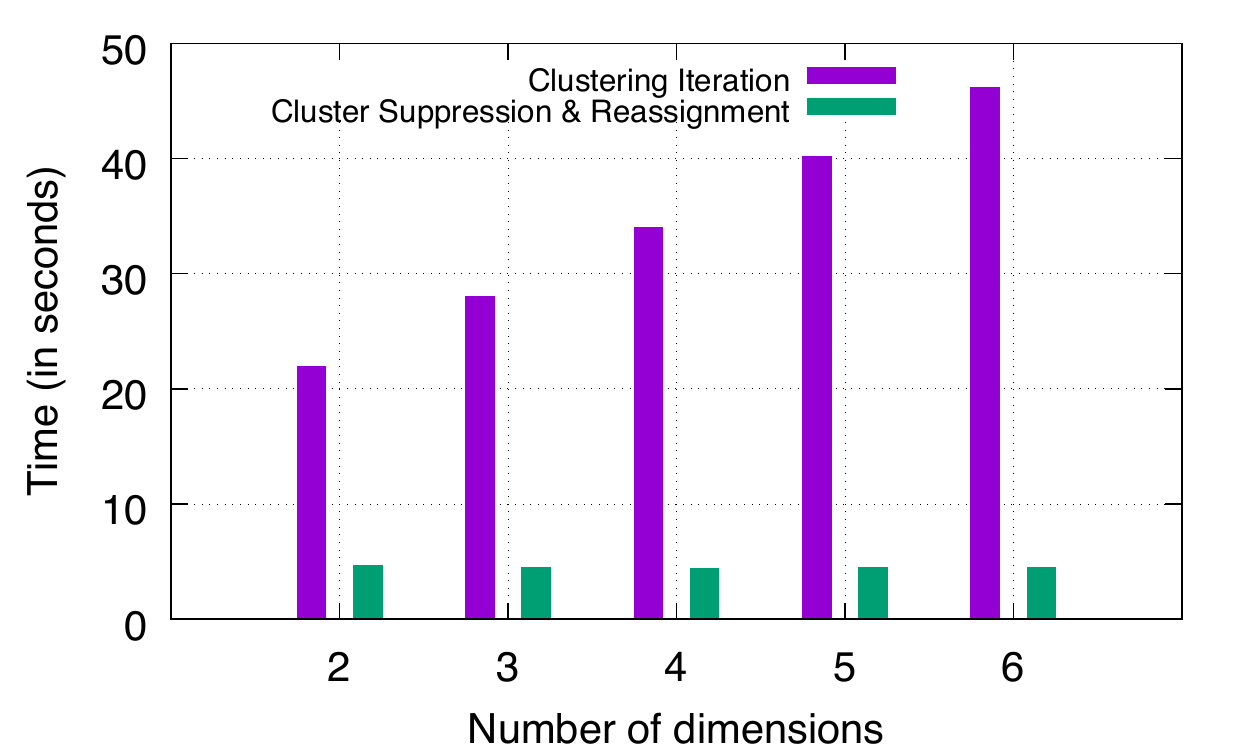}}

	\caption{Execution time for varying number of dimensions with 1800 data points}
	\label{fig:vary-dim}
\end{figure*}

\item {\bf RecomputeClusterCentres : } In this phase, Party \CB gains access (by virtue of decryption) to the following plaintext (but randomized) quantities:
\begin{itemize}
 \item Sum of data points nearest in each cluster centre
 \item Number of data points in each cluster centre
 \end{itemize}
 We note that since these quantities are multiplicatively randomized by Party \CA, their actual values are effectively hidden from Party \CB. It is also worth noting that the randomization used is different for each cluster, implying that Party \CB cannot hope to leverage any sharing\slash re-use of randomization across different cluster centres to gain additional information about the sum or number of data points for any given cluster centre. 

\item {\bf Non-kClusters : } In Algorithm \ref{alg:nonk}, Party \CB gets access to the plaintext (but masked) of the anonymization factor $k$ and the number of data points in each cluster center. But since Party \CA evaluates a random polynomial \emph{poly(x)} over their encrypted values before sending them, hence Party \CB does not learn the actual anonymization factor $k$ and the number of data points in each cluster centre. A similar proof as shown in Step \ref{sec:pb:CompluteMinIndex} above can be presented here.

\item {\bf Suppress and Reassign Clusters : } In this step, Party \CB receives a permuted vector of size $k'$ having the encrypted counts of the number of elements in non-$k$ clusters padded with some fake values. Hence after decryption of this vector Party \CB cannot identify the number of elements in the non-$k$ clusters, since we have picked a secure pseudo-random
permutation, which is computationally difficult to invert, implying that the exact identity of cluster centres associated with any cluster count is hidden from Party \CB.

\end{enumerate}

The above leakage profile for Party \CB leads to the following security guarantee :

\begin{Theorem}{\textnormal{\textbf{Security Guarantee for Party \CB:}}}
\emph{The secure $k$-Anonymization protocol leaks no information to Party \CB except that it only learns if an attribute is a direct identifier but does not gain any knowledge about the encrypted data points or the cluster to which a data point is assigned and the cluster centre representatives.}
\end{Theorem}

\section{Performance}
\label{sec:performance}

In this section, we empirically evaluate the performance of our protocols.
The experimental setup consists of three machines, representing the Data Owner, Party \CA and Party \CB. The configuration of machines representing Party \CA and Party \CB is: 4 core 2.8 GHz processors, 64 GB RAM running Ubuntu 16.04 LTS; the configuration of the machine representing Data Owner is: 4 core 2.8 GHz processors, 8 GB RAM running Ubuntu 16.04 LTS. We use the HELib~\cite{HELib} library to encrypt the data using LFHE. Specifically, for HELib we set (i) $p = 1099511627689$, a large prime between $2^{40}$ and $2^{87}$, (ii) the maximum depth to $10$ and (iii) the security parameter to $128$. 

The two parameters affecting the performance of our protocols are the number of data points and the number of dimensions in the data. To study the independent effect of each of these parameters on our protocols, we use simulated data. We generated two datasets, one with a varying number of data points (results shown in Figure~\ref{fig:vary-data-points}) and one with a varying number of dimensions (results shown in Figure~\ref{fig:vary-dim}). The data were generated using a uniform distribution. We repeated each experiment multiple times with a newly generated dataset. The average time across these experiments is reported here.

LFHE allows SIMD operations by packing multiple plaintext data values into a single structure and then encrypting them together into a single ciphertext. We utilize this feature of LFHE extensively. We encrypt each dimension of the data point independently. For each dimension, we pack data from multiple data points into a single structure and then encrypt this structure to get a single ciphertext. This ciphertext is then outsourced to Party~\CA. The time taken to encrypt the plaintext data is shown in Figure \ref{fig:dp-encrypt} and Figure \ref{fig:dim-encrypt}. These figures clearly show that the data encryption time scales linearly with the number of data points and the number of dimensions.

The second major step in our protocols is to check if a particular combination of dimensions is a privacy vulnerability identifier or not. The actual number of combinations that need to be tested is data-dependent. To remove this data dependence from the performance evaluation, we report the average time taken to identify a quasi-identifier. The results are shown in Figure \ref{fig:dp-check-identifier} and Figure \ref{fig:dim-check-identifier}. From the figures, it is clear that the time taken to identify a quasi-identifier scales linearly with the number of data points and is independent of the number of dimensions (this is because, the most computationally heavy step is decryption of distance at Party~\CB, which is independent of the number of dimensions).

Once a quasi-identifier is identified, the next step is to cluster the data in the quasi-identifiers. Furthermore, after clustering, we use the ``Cluster to Cluster'' re-assignment strategy to eliminate non-$k$ clusters. Both of these operations are highly dependent on the data and the choice of initial cluster centres. To remove this data dependence from the performance evaluation we report the average time taken for each iteration of clustering and the time taken to re-assign a single cluster. Figure~\ref{fig:dp-clustering-per-iteration} and Figure~\ref{fig:dim-clustering-per-iteration} show that both the above operations scale linearly with the number of data points. The number of dimensions has a negligible effect on the cluster reassignment (again, the most expensive step being decryption of inter-cluster distance at Party~\CB, which is independent of the number of dimensions).

The above performance evaluation shows that our protocols scale linearly with the number of data point as well as the number of dimensions in the dataset.

\section{Conclusions}
\label{sec:conclusions}

This paper presents a set of secure algorithms on how to apply anonymization over homomorphically encrypted databases.
It does not focus on a single anonymization approach but touches various components that are required for end-to-end privacy.
It demonstrated how to achieve uniqueness discovery, data masking, differential privacy and $k$-anonymity over encrypted data without leaking information about original values. 
Feasibility of this solution is shown by empirical evaluation.
This work is the first to perform several techniques, like vulnerability assessment, differential privacy and $k$-anonymity, 
over encrypted datasets which means there is room for improvement and future work, especially on the performance and optimization side.


\bibliographystyle{IEEEtran}
\bibliography{paper}

\appendix
\section{Leakage from ordered equations}
\label{sec:ordeq}
We now examine the possibility of any leakage to Party~\CB from the resulting system of ordered equations. Let $d^{''}_{i,1} < d^{''}_{i,2} < \cdots < d^{''}_{i,k}$ be the ordered set of plaintext distances, and $poly(d^{''}_{i,1}) < poly(d^{''}_{i,2}) < \cdots < poly(d^{''}_{i,k})$ be the ordered set of polynomial outputs obtained by Party~\CB upon decryption. As mentioned earlier, the polynomial $poly(x)$ is of the form $a_0+a_1\cdot x +a_2\cdot x^2 + \cdots + a_p\cdot x^p$ for some random $p\in\mathbb{N}$. Party~\CB can formulate the following system of equations for $j \in \{1,k'\}$:
\begin{equation}
\hspace*{25pt} poly(d^{''}_{i,j}) = a_0+a_1\cdot d^{''}_{i,j} +a_2\cdot (d^{''}_{i,j})^2 + \cdots + a_p\cdot (d^{''}_{i,j})^p\nonumber
\end{equation}
\noindent \noindent where only the left hand side of each equation is known to Party~\CB. Without loss of generality, we may assume that Party~\CB can guess with high probability the degree $p$ of the polynomial chosen by Party~\CA, as well as the range of values (say $[0,2^N]$) that each plaintext distance $d^{''}_i$ can take. This is a particularly relevant assumption in the context of real world datasets, where the adversary may possess some apriori knowledge of the range of Euclidean distances between the data points. In addition, since homomorphic polynomial evaluation in the encrypted domain is a costly operation, the degree $p$ can only take a small range of values, which Party~\CB can also accurately guess in a small number of trials. However, we prove that even if Party~\CB has full knowledge of the aforementioned parameters, it cannot recover the original data points within a feasible amount of computation time. Observe that the system of equations has exactly $k'+p+1$ unknown variables from Party~\CB's point of view, while the number of equations is only $k$. Hence, Party~\CB must correctly guess the $p+1$ smallest distances $d^{''}_{i,1}, d^{''}_{i,2},\cdots,d^{''}_{i,p+1}$ to recover the polynomial coefficients. The average number of possible values that these distances can take is $\binom{2^N}{p+1}$, which is approximately the same as $2^{N\cdot (p+1)}$ for $2^N \gg (p+1)$. In other words, the probability that Party~\CB successfully recovers the polynomial coefficients, and subsequently the plaintext distances, is approximately $1/2^{N\cdot (p+1)}$, which is close to negligible. For example, for $N=16$ and $p=9$, the probability that Party~\CB is able to recover the plaintext distances is approximately $2^{-160}$, which is close to negligible for a security level of $160$ bits. Thus, even when the range of plaintext distances and the degree of the polynomial chosen by Party~\CA are reasonable small and known apriori to Party~\CB, the information leakage is negligible. Also note that Party~\CA refreshes the polynomial for each data point, implying that Party~\CB gains no additional information across the data points. Finally, even if Party~\CB is able to recover the plaintext distances in some extreme cases (e.g., when the plaintext distance values follow some specific pattern), it still does not directly reveal the plaintext data points to Party~\CB, as the candidate cluster centres are also unknown. 

We clarify here that the ordered set of polynomial outputs obtained by Party~\CB upon decryption are not necessarily uniformly distributed over the entire plaintext space since this would potentially lead to \emph{wrap-arounds} and make it impossible to preserve ordering, which is crucial to the correctness of the protocol. It turns out that, for our security argument to hold, the distribution of the polynomial outputs need not be uniformly random. Recall that the probability that Party~\CB is able to recover the plaintext distances is bounded from above by $1/2^{N\cdot (p+1)}$, where $N$ is the maximum plaintext distance value pertaining to a given data set. Hence, so long as the semi-honest Party~\CA chooses $p$ to be sufficiently large, recovering the plaintext distances is computationally infeasible for Party~\CB.

\end{document}